\documentclass[aps,pra,reprint,amsmath,amssymb,superscriptaddress,floatfix]{revtex4-2}
\usepackage{bm}
\usepackage{graphicx}
\usepackage{booktabs}
\usepackage{mathtools}
\usepackage{microtype}
\usepackage{amsthm}
\usepackage[hidelinks]{hyperref}
\hypersetup{pdftitle={Pump-scheme dependence of few-emitter lasers: exact multilevel correlations and photon statistics},pdfauthor={Stephan Hartmann}}
\newtheorem{proposition}{Proposition}

\newtheorem{lemma}{Lemma}
\newtheorem{remark}{Remark}
\newcommand{\codeDOI}{10.5281/zenodo.23841957}
\newcommand{\Tr}{\operatorname{tr}}
\newcommand{\Exc}{\mathcal E}
\graphicspath{{figures/}}
\newcommand{\mv}{\boldsymbol m}
\newcommand{\nop}{\hat n}

\begin{document}

\title{Pump-scheme dependence of few-emitter lasers:\texorpdfstring{\\}{ }
exact multilevel correlations and photon statistics}

\author{Stephan Hartmann}
\affiliation{Munich Center for Mathematical Philosophy, LMU Munich, Germany}

\date{\today}

\begin{abstract}
Few-emitter laser models usually replace the excitation cycle of the
gain medium by a single Markovian pump jump of a two-level emitter.  We
test this replacement with the cavity and laser-transition parameters
held fixed and the two-level pump rate fitted to the same mean photon
number, using an exact damping-basis continued-fraction solution of the
cavity--emitter master equation for \(N\) identical \(d\)-level emitters
with arbitrary independent incoherent transitions and one
cavity-coupled transition, in the known reduced space of level
populations and lasing coherences.  First, a pump-quantum filtration
shows that the inter-emitter coherence of \(\Lambda\) and ladder media
starts at second rather than first order in the pump; we evaluate its
leading coefficient from the exact finite-system series and map the
recycling rate at which its sign changes.  Second, at matched photon
number, rate-limited steps in the pump cycle push the long-time
output-counting Fano factor of two and three emitters well below one:
for two emitters it reaches \(0.66\) at a mean cavity occupation of
\(0.3\) and is still \(0.74\) at an occupation of one, where
\(g^{(2)}(0)\) is within \(3\%\) of unity, because the anticorrelation
sits at delays of about half a cavity lifetime rather than at zero
delay.  For two and three emitters the fitted two-level model gets both
statistics wrong and places the output above the shot-noise level.
Against independent single-emitter lasers at the same output,
the common cavity erodes rather than enhances this single-emitter
effect.
\end{abstract}

\maketitle

\section{Introduction}
\label{sec:intro}

Few-emitter laser theory usually models the gain medium as a two-level
emitter with an incoherent pump term \(P\,\mathcal D[\tau_+]\)
\cite{MuSavage1992,Ginzel1993,RiceCarmichael1994,Loffler1997,Kilin2002,MeiserHolland2010,KirtonKeeling2018}.
This is an effective description.  Real gain media pump through
auxiliary levels: a \(\Lambda\) scheme empties the lower laser level into
a reservoir that is pumped into the upper laser level, a four-level
ladder separates the pump and laser transitions entirely, and a V scheme
pumps the lower laser level into a fast-decaying upper reservoir.  The
distinction has consequences that the effective pump rate cannot
capture.  The pump acts as a dephasing channel for the lasing coherence
only when it empties a laser level, so that the two-level model carries
a pump-induced broadening that \(\Lambda\) and ladder media do not have;
a pump cycle with several rate-limited steps delivers excitations more
regularly than a Poisson process and can drive the output below the
shot-noise level --- the pump-noise suppression of regularly pumped
lasers \cite{Yamamoto1986,Haake1989} in the dynamic form it takes in
multilevel lasers \cite{Ritsch1991,RitschZoller1992,Briegel1996}; and the number of pump quanta an
emitter must absorb before it can share a coherence with another emitter
sets the order at which inter-emitter correlations appear.  Single-ion
lasers \cite{Dubin2010}, atoms in optical cavities \cite{McKeever2003},
and quantum-dot nanolasers \cite{Jahnke2016,Ojambati2024} operate with
such multilevel kinetics at the level of a few emitters and a few
photons, where cumulant closures are least controlled
\cite{Gies2007,Leymann2014} and exact finite-\(N\) solutions are the
natural benchmark.

Exact permutation-invariant treatments of many multilevel emitters exist
\cite{ChaseGeremia2008,XuTieriHolland2013,Hartmann2016,GeggRichter2016,Shammah2018}.
In particular, Gegg and Richter \cite{GeggRichter2016} showed that for
incoherently pumped three- and four-level lasers the optical coherences
that the dynamics never generates decouple from the permutation-invariant
equations of motion, which reduces the scaling of the number of
variables from \(N^8\) to \(N^4\) and from \(N^{15}\) to \(N^5\), and they
solved the resulting Fock-space problems numerically.  In a companion
paper \cite{PaperI}, hereafter Paper~I, we combined the Briegel--Englert
damping basis of the cavity mode \cite{BriegelEnglert1993,EnglertMorigi2002}
with the permutation-invariant Liouville basis of \(N\) two-level
emitters and showed that the coupled Liouvillian is an exact block-Jacobi
operator in a single radial field index, whose stationary state is a
matrix continued fraction with a linear, uniquely solvable asymptotic
closure, and whose convergence to the stationary state of the
infinite-dimensional master equation could be proved for a
zero-temperature cavity under explicit hypotheses.  The present paper
carries that construction over to \(N\) identical \(d\)-level emitters
with arbitrary independent incoherent level kinetics and one
cavity-coupled transition, and uses it to answer the physical questions
above.  Methodologically it contributes a formulation of the reduced operator
space for arbitrary incoherent rate graphs, generalizing the three- and
four-level cases of Ref.~\cite{GeggRichter2016}
(Proposition~\ref{prop:invariant}), a single lifting rule that replaces
scheme-specific multiplication tables (Lemma~\ref{lem:lift}), a
transfer of the structural and convergence results of Paper~I with the
hypotheses under which they hold (Proposition~\ref{prop:carry}), a
self-contained assembly prescription for the recurrence
(Appendix~\ref{app:assembly}) with a stand-alone reference
implementation, and a normalized-coordinate implementation that
improves the conditioning of the recursion
(Appendix~\ref{app:normalized}).  Spectra
and intensity correlations can be obtained from Fock-space solvers by
quantum regression as well; what the damping basis adds is that they,
the moment functionals, and a weak-pump series that terminates exactly
in the radial index at each order all follow from one recurrence with a
controlled structure.

Whether an effective two-level pump ``fails'' depends on what is held
fixed and what is fitted, because a two-level model with more free
parameters can be tuned to match more observables.  We ask what a single Markovian pump jump
gets wrong when the cavity (\(A\), \(\nu\), \(\Delta\)), the coupling
\(g\), the laser-transition decay \(\gamma\), and \(N\) are those of the
multilevel medium and only the pump rate is fitted, to the mean photon
number.  This is the replacement that effective-pump models make; fits
with additional free parameters (an extra dephasing, a renormalized
\(\gamma\) or \(g\)) are not tested here.  The physical results are the
following.  (i) We prove that, for a
zero-temperature cavity, the photon number of any scheme whose pump
empties a laser level tends to zero as the pump rate grows
(Proposition~\ref{prop:quench}) --- a sufficient mechanism for
self-quenching, not a classification --- and identify the explicit
limiting models to which the \(\Lambda\) and ladder cycles studied here
saturate instead; the exact solution follows the two-level and V lasers
through self-quenching and the \(\Lambda\) and ladder lasers through
saturation, and shows that matching the absorbed pump current reduces
but does not remove the differences between schemes.  (ii) We
prove a pump-quantum filtration (Proposition~\ref{prop:pumpquantum}) by
which the pair coherence of \(\Lambda\) and ladder media is of second
order in the pump, evaluate its leading coefficient from a finite
linear system that is exact for the model, and map the recycling-rate
boundary at which this coefficient changes sign.  (iii) We show, by a
controlled comparison at matched photon number, that rate-limited steps
in the pump cycle lower the long-time output-counting Fano factor of
two and three emitters below one, and the intracavity Fano factor as
well at photon numbers below one; at a mean occupation of one the
counting Fano factor of two emitters is \(0.74\) while
\(g^{(2)}(0)=0.97\), and the time-resolved intensity correlation shows
why: the anticorrelation is concentrated at delays of about half a
cavity lifetime.  The fitted two-level model gets \(g^{(2)}(0)\), the
intracavity and counting Fano factors, and the weak-pump order and sign
of the pair coherence wrong; a reference of independent single-emitter
lasers at the same output quantifies what the common cavity changes;
and an independent Fock-space calculation confirms the counting
statistics to \(10^{-12}\).  (iv) We decompose the output flux exactly
over the collective sectors and quantify the population of the
structurally dark sectors that exist for every \(N\) when \(d\ge3\).
(v) We benchmark the second-order cumulant closure for multilevel media
to \(N=8\).  Neither pump-noise suppression nor exact permutation-invariant
multilevel calculations are new; what is new here is their combination:
several correlated emitters treated exactly with their full pump cycle,
compared at matched intensity and against an independent-source
reference, together with analytic selection rules for the pump order of
the inter-emitter coherence and an exact-series determination of its
recycling-dependent sign, and the demonstration that the
temporal noise suppression survives where the equal-time statistics
obscure it.

Section~\ref{sec:model} fixes the model, Sec.~\ref{sec:method} gives
the exact solution in compact form, Secs.~\ref{sec:weakpump}--\ref{sec:Nscaling}
contain the physical results, and Sec.~\ref{sec:discussion} the scope
and limitations.  Appendix~\ref{app:construction} contains the proofs
and construction details of Sec.~\ref{sec:method},
Appendix~\ref{app:carry} records which properties of
the emitters each inherited proof uses, Appendix~\ref{app:closure} gives
the multilevel closure equations, Appendix~\ref{app:validation} the
validation, cost, and conditioning, Appendix~\ref{app:normalized}
the solver used for the largest emitter numbers, and
Appendix~\ref{app:assembly} the explicit assembly of the recurrence.
All code and data are available \cite{code}.

\section{Model and pump schemes}
\label{sec:model}

Each of \(N\) identical emitters has levels \(|1\rangle,\ldots,|d\rangle\)
with projectors \(P_i=|i\rangle\langle i|\).  Two of them, the upper and
lower laser levels \(u\) and \(l\), are coupled to one cavity mode
through \(\tau_+=|u\rangle\langle l|\), \(\tau_-=\tau_+^\dagger\), and the
collective operators \(S_\pm=\sum_j\tau_\pm^{(j)}\).  Incoherent
transitions \(i\to i'\) occur at rates \(\gamma_{i'\leftarrow i}\ge0\),
and the lasing coherence may suffer pure dephasing at rate
\(\gamma_\phi\).  With \(\Gamma_i=\sum_{i'\neq i}\gamma_{i'\leftarrow i}\)
the escape rate from level \(i\), \(\nop=a^\dagger a\), and
\(\mathcal D[L]\rho=L\rho L^\dagger-\tfrac12\{L^\dagger L,\rho\}\), the
master equation is
\begin{align}
\dot\rho={}&-\frac{i}{\hbar}[H,\rho]+A(1+\nu)\mathcal D[a]\rho+A\nu\mathcal D[a^\dagger]\rho\notag\\
&+\sum_{j=1}^N\Big[\sum_{i\neq i'}\gamma_{i'\leftarrow i}\,\mathcal D\big[|i'\rangle\langle i|^{(j)}\big]\rho
+2\gamma_\phi\,\mathcal D\big[P_u^{(j)}\big]\rho\Big],
\label{eq:master}\\
H={}&\hbar\omega\nop+\hbar\sum_j\sum_i\epsilon_iP_i^{(j)}-\frac{\hbar g}{2}\big(a^\dagger S_-+aS_+\big),
\label{eq:H}
\end{align}
with cavity decay rate \(A\), thermal occupation \(\nu\), transition
frequency \(\Omega=\epsilon_u-\epsilon_l\), and detuning
\(\Delta=\Omega-\omega\).  The two-level model of Paper~I is the case
\(d=2\), \(\gamma_{l\leftarrow u}=B(1-s)\), \(\gamma_{u\leftarrow l}=Bs\),
\(\gamma_\phi=C-B/2\).  Throughout, \(A>0\), \(g\neq0\), and the rate
graph is irreducible on the \(d\) levels, so that the classical generator
\begin{equation}
W_{i'i}=\gamma_{i'\leftarrow i}\ (i'\neq i),\qquad W_{ii}=-\Gamma_i,
\label{eq:W}
\end{equation}
has a unique stationary distribution \(\bar p\).  Every Fock-truncated
Lindbladian is then irreducible by the argument of Paper~I and has a
unique stationary state \cite{Spohn1977,Frigerio1978}.  For the
untruncated equation at \(\nu=0\) we keep the moment-stationarity
uniqueness hypothesis (U) of Paper~I with the excitation operator
\(\Exc=\nop+N_u\), \(N_u=\sum_jP_u^{(j)}\): every trace-class \(\sigma\)
with \(e^{\theta\Exc}\sigma e^{\theta\Exc}\) trace class for all
\(\theta>0\) and \(\Tr(\sigma\mathcal L^\dagger Y)=0\) for all polynomials
\(Y\) in the field and emitter operators is a multiple of one density
operator \(\rho_{\rm ss}\).  The Lyapunov inequality that makes (U)
plausible in Paper~I holds with
\(\mathcal L^\dagger\Exc\le-\min(A,\Gamma_u)\Exc+\Gamma_\uparrow N+A\nu\),
\(\Gamma_\uparrow=\max_{i\neq u}\gamma_{u\leftarrow i}\), provided
\(\Gamma_u>0\).

\begin{figure}[tbp]
\includegraphics[width=\linewidth]{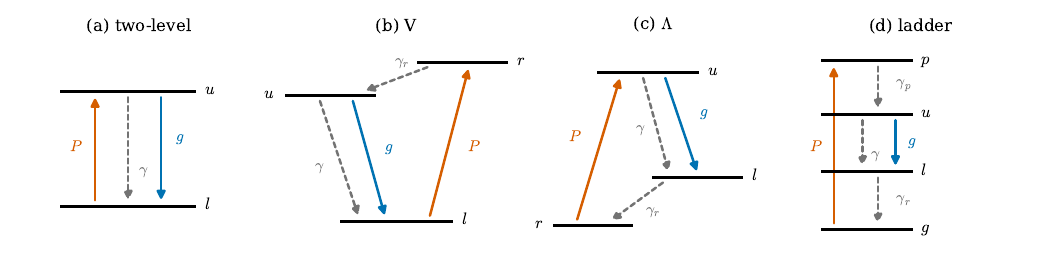}
\caption{The four pump schemes.  Solid blue: cavity-coupled laser
transition (coupling \(g\)); orange: incoherent pump (rate \(P\));
dashed grey: incoherent decays.  (a) Two-level emitter: the pump empties
the lower laser level into the upper one.  (b) V scheme: the lower laser
level is pumped into a reservoir \(r\) that decays at rate \(\gamma_r\)
into the upper laser level.  (c) \(\Lambda\) scheme: a reservoir below
the laser levels is pumped into the upper laser level; the lower laser
level is emptied into the reservoir at the recycling rate \(\gamma_r\).
(d) Four-level ladder: ground \(\to\) pump level (\(P\)) \(\to\) upper
laser level (\(\gamma_p\)) \(\to\) lower laser level (cavity and
\(\gamma\)) \(\to\) ground (\(\gamma_r\)).  In all schemes the laser
transition also decays spontaneously at rate \(\gamma=\gamma_{l\leftarrow u}\).}
\label{fig:schemes}
\end{figure}

Three schemes serve as examples throughout (Fig.~\ref{fig:schemes},
Table~\ref{tab:schemes}): the V scheme and the \(\Lambda\) scheme
(\(d=3\)) and the four-level ladder (\(d=4\)).  What distinguishes them
is which level the pump empties.  In the two-level and V schemes it
empties the lower laser level; in the \(\Lambda\) scheme and the ladder
it empties a reservoir, and neither laser level is emptied by the pump
--- the lower laser level is emptied by recycling instead.  All rates
are quoted in units of the cavity decay rate \(A\).  Two parameter
regimes are used: the good-cavity regime of Paper~I, \(g=5A\),
\(\gamma=A\) (its decay rate \(B(1-s)\) at \(B=10A\), \(s=0.9\); the
two-level laser at \(P=9A\) is then the good-cavity laser of Paper~I and
reproduces its Table~VIII for \(N=1\)--\(3\)), in which the lasers of Sec.~\ref{sec:strongpump} reach
several photons, and the bad-cavity regime \(g=1.1A\), \(\gamma=0.3A\),
in which the few-quanta correlation results of Secs.~\ref{sec:weakpump},
\ref{sec:dark}, and \ref{sec:Nscaling} are obtained.

\begin{table*}[tbp]
\caption{The example schemes.  \(C_{ul}\) is the transverse damping rate
of the lasing coherence, Eq.~(\ref{eq:Cul}), at \(\gamma_\phi=0\);
\(D_N\) the atomic block dimension (\(L=d+2\) local operators,
Proposition~\ref{prop:invariant}); \(\chi\) the pump-quantum charge of
Proposition~\ref{prop:pumpquantum} that gives the sharpest weak-pump
orders, with the resulting order of the pair coherence; and the
\(P\to\infty\) limit of the scheme at \(\nu=0\) (Sec.~\ref{sec:strongpump}).}
\label{tab:schemes}
\begin{ruledtabular}
\begin{tabular}{lclccll}
scheme & \(d\) & transitions (rate) & \(C_{ul}\) & \(D_N\) & \(\chi\) (\(C_2\) order) & \(P\to\infty\) limit\\
\hline
two-level & 2 & \(l\!\to\!u\ (P)\), \(u\!\to\!l\ (\gamma)\) & \(\frac{\gamma+P}{2}\) & \(\binom{N+3}{3}\) & \(\chi_l=0,\chi_u=1\) (\(P\)) & \(\langle\nop\rangle\to0\)\\
V & 3 & \(l\!\to\!r\ (P)\), \(r\!\to\!u\ (\gamma_r)\), \(u\!\to\!l\ (\gamma)\) & \(\frac{\gamma+P}{2}\) & \(\binom{N+4}{4}\) & \(\chi_l=0,\chi_r=\chi_u=1\) (\(P\)) & \(\langle\nop\rangle\to0\)\\
\(\Lambda\) & 3 & \(r\!\to\!u\ (P)\), \(u\!\to\!l\ (\gamma)\), \(l\!\to\!r\ (\gamma_r)\) & \(\frac{\gamma+\gamma_r}{2}\) & \(\binom{N+4}{4}\) & \(\chi_r=0,\chi_u=\chi_l=1\) (\(P^2\)) & two-level, pump \(\gamma_r\)\\
ladder & 4 & \(g\!\to\!p\ (P)\), \(p\!\to\!u\ (\gamma_p)\), \(u\!\to\!l\ (\gamma)\), \(l\!\to\!g\ (\gamma_r)\) & \(\frac{\gamma+\gamma_r}{2}\) & \(\binom{N+5}{5}\) & \(\chi_g=0,\chi_{p,u,l}=1\) (\(P^2\)) & V, pump \(\gamma_r\), reservoir \(\gamma_p\)\\
\end{tabular}
\end{ruledtabular}
\end{table*}
\section{Exact solution}
\label{sec:method}

This section states the construction; proofs and the details of the
local bases are in Appendix~\ref{app:construction}, and the reader
interested in the physics may skip to Sec.~\ref{sec:weakpump} after
Eq.~(\ref{eq:flux}).

\subsection{Reduced local operator space and collective basis}

With one cavity-coupled transition and incoherent everything else, the
dynamics of one emitter never leaves the \((d+2)\)-dimensional space
\(\mathcal A=\operatorname{span}\{P_1,\ldots,P_d,\tau_+,\tau_-\}\): jumps
map populations to populations and only damp the coherence they touch,
and the cavity generates no coherence except \(|u\rangle\langle l|\) and
its conjugate.  For incoherently pumped three- and four-level lasers this
decoupling of the unused coherences was demonstrated and exploited by
Gegg and Richter \cite{GeggRichter2016}; the following proposition states
it for arbitrary independent incoherent rate graphs.

\begin{proposition}[Invariant local operator space]
\label{prop:invariant}
(i) The one-emitter Liouvillian of Eq.~(\ref{eq:master}), including the
level Hamiltonian, maps \(\mathcal A\) into itself; (ii) left and right
multiplication by \(\tau_\pm\) map \(\mathcal A\) into itself; (iii)
\(\openone\in\mathcal A\).  Consequently the span of the field
damping-basis operators tensored with the permutation-invariant products
of elements of \(\mathcal A\) is invariant under the full Liouvillian,
and the stationary state of every Fock truncation, as well as
\(\rho_{\rm ss}\) under (U), lies in it.  The atomic block dimension is
\begin{equation}
D_N^{(d)}=\binom{N+d+1}{d+1},
\label{eq:DN}
\end{equation}
\(\binom{N+4}{4}\) for three and \(\binom{N+5}{5}\) for four levels,
against \(\binom{N+d^2-1}{d^2-1}\) for the full local Liouville space
(Table~\ref{tab:dims}).
\end{proposition}

A basis \(\{b_\mu\}\) of \(\mathcal A\) with exactly one traceful element
\(b_0\), traceless \(b_1,\ldots,b_{d-1}\), and \(b_\pm=\tau_\pm\) is
chosen; we use the eigenoperators of the one-emitter Liouvillian (the
local damping basis) where its rate matrix is diagonalizable and a
rate-independent fixed basis otherwise (Appendix~\ref{app:construction}).
In either, \(\tau_\pm\) are eigenoperators with eigenvalues
\(-C_{ul}\mp i\Omega\),
\begin{equation}
C_{ul}=\tfrac12(\Gamma_u+\Gamma_l)+\gamma_\phi ,
\label{eq:Cul}
\end{equation}
the transverse rate of the lasing coherence: half the sum of the escape
rates of the two laser levels plus pure dephasing.  For the two-level
model this is \(C=B/2+\gamma_\phi\); for the V scheme
\(C_{ul}=(\gamma+P)/2\) grows with the pump exactly as in the two-level
model; for the \(\Lambda\) scheme and the ladder \(C_{ul}=(\gamma+\gamma_r)/2\)
is pump independent.  The cavity is expanded in the damping basis of
Briegel and Englert \cite{BriegelEnglert1993}, the right eigenoperators
\(\rho_n^{(k)}\) of the field Liouvillian with radial index \(n\ge0\),
coherence grade \(k\in\mathbb Z\), eigenvalues \(-A(n+|k|/2)-i\omega k\),
and \(\Tr\rho_n^{(k)}=\delta_{n0}\delta_{k0}\).  The permutation-invariant
atomic basis consists of the averaged symmetrized products
\begin{equation}
R_{\mv}=\frac{1}{\mathcal N(\mv)}\sum_{\pi}^{\rm distinct}\pi\Big[\bigotimes_\mu b_\mu^{\otimes m_\mu}\Big],\quad
\mathcal N(\mv)=\frac{N!}{\prod_\mu m_\mu!},
\label{eq:Rm}
\end{equation}
with \(\mv=(m_0,\ldots,m_{d-1},m_+,m_-)\), \(|\mv|=N\),
\(\Tr R_{\mv}=\delta_{\mv,Ne_0}\), and coherence charge
\(q(\mv)=m_+-m_-\).  The action of any one-site map on this basis follows
from one rule, which replaces the four multiplication tables of Paper~I.

\begin{lemma}[Lifting rule]
\label{lem:lift}
Let \(x\) be a linear map on \(\mathcal A\) with structure constants
\(x\,b_\mu=\sum_\nu x_{\nu\mu}b_\nu\), and \(x^{(N)}=\sum_jx^{(j)}\).  Then
\begin{equation}
x^{(N)}R_{\mv}=\sum_\mu m_\mu\sum_\nu x_{\nu\mu}\,R_{\mv-e_\mu+e_\nu}.
\label{eq:lift}
\end{equation}
\end{lemma}

\subsection{Block-Jacobi recurrence and what carries over}
\label{sec:cf}

The total coherence \(K=k+q(\mv)\) of a product
\(\rho_n^{(k)}\otimes R_{\mv}\) is conserved by the interaction.  With an
ordering \(\{\mv_\alpha\}\), \(k_\alpha=K-q_\alpha\),
\(\Phi_{n\alpha}^{(K)}=\rho_n^{(k_\alpha)}\otimes R_{\mv_\alpha}\), and
\(\rho^{(K)}=\sum_{n,\alpha}(X_n)_\alpha\Phi^{(K)}_{n\alpha}\), the lifted
matrices of left and right multiplication by \(\tau_\pm\) and the field
coefficients of the damping basis give blocks
\(M_n,G_n,F_n\in\mathbb C^{D_N\times D_N}\), Eq.~(\ref{eq:blocks}) of
Appendix~\ref{app:assembly}, which lists every ingredient needed to
implement them, with the diagonal uncoupled part
\(-A(n+|k_\beta|/2)+\lambda^{\rm at}_\beta-i\omega K-i\Delta q_\beta\),
where \(\lambda^{\rm at}_\beta=-\sum_\kappa m_{\kappa\beta}\beta_\kappa-C_{ul}(m_{+\beta}+m_{-\beta})\)
in the damping basis (in the fixed basis the atomic matrix is not
diagonal, which nothing below requires).  Call \(|k_\beta|\) the grade of a basis state; it ranges over
\(0,\ldots,N+|K|\), because \(|q|\le N\) for one cavity-coupled
transition.

\begin{proposition}[Structure of the \(d\)-level recurrence]
\label{prop:carry}
For \(N\) identical \(d\)-level emitters obeying Eq.~(\ref{eq:master})
and any fixed \(K\):
(i) the Liouvillian on the permutation-invariant sector is exactly block
tridiagonal in the radial index,
\(\dot X_n=M_nX_n+G_nX_{n+1}+F_nX_{n-1}\), \(X_{-1}=0\);
(ii) \(G_n=G\) is constant, \(F_n=nF^{(1)}\), \(M_n=nM^{(1)}+M^{(0)}\)
with \(M^{(1)}=-A\openone+\tfrac{ig}{2}\mathcal N_K\),
\(\mathcal N_K=[S_+,\cdot]\mathcal P_{q<K}+[S_-,\cdot]\mathcal P_{q>K}\),
so that a finite limit of the transfer matrices \(X_{n+1}=R_nX_n\)
satisfies the linear equation \(M^{(1)}R_\infty=-F^{(1)}\), uniquely
solvable for every \(A>0\) by the finite Neumann series
\(R_\infty=A^{-1}\sum_{j=0}^{N+|K|}(ig/2A)^j\mathcal N_K^{\,j}F^{(1)}\);
(iii) \(F^{(1)}\) and \(\mathcal N_K\) lower the grade of every basis
state by exactly one, \(G\) raises it by exactly one, and \(R_\infty\)
lowers it by at least one; all four are nilpotent of index at most
\(N+|K|+1\);
(iv) for every terminal matrix \(R_{n_{\max}}\) in a ball
\(\|R\|\le\rho_0\), \(\rho_0>\|R_\infty\|\), the backward recursion
\(R_n=-(M_{n+1}+GR_{n+1})^{-1}F_{n+1}\) converges as \(n_{\max}\to\infty\),
for all \(n\ge n_0(\rho_0)\), to limits \(R_n^\infty=R_\infty+O(1/n)\)
independent of the terminal matrix; the same holds for all \(n\ge0\)
under
\begin{itemize}
\item[(P)] the limiting pivots \(M_j+GR_j^\infty\) are invertible for
\(1\le j\le n_0\) (for \(j>n_0\) this is automatic);
\end{itemize}
(v) at \(\nu=0\) the stationary state of every Fock truncation obeys
\(\langle e^{\theta\Exc}\rangle\le2e^{\theta(N+n_*)}\),
\(n_*=\lceil2\Gamma_\uparrow N/(Ae^{-\theta})\rceil\), uniformly in the
cutoff, the Fock truncations converge in trace norm to
\(\rho_{\rm ss}\) under (U), and for \(K=0\), under (U) and (P), the
normalized solution of the continued fraction converges componentwise
to the damping-basis coefficients of \(\rho_{\rm ss}\) and every
observable functional below to its stationary value.
\end{proposition}

Items (i)--(iv) are Propositions~1--4 of Paper~I and item (v) is its
Lemma~1, Proposition~5, and Theorem~1; Appendix~\ref{app:carry} lists,
item by item, the only properties of the emitter sector that the
original proofs use --- essentially that \(\tau_+\) raises the coherence
charge by exactly one on every nonvanishing term of
Eq.~(\ref{eq:lift}), that \(|q|\le N\), and that the jumps change
\(\Exc\) by \(0\) or \(\pm1\) with bounded raising rate --- and verifies
that each holds here.  As in Paper~I, (P) concerns the finitely many
non-asymptotic limiting pivots and remains an assumption whose
finite-cutoff proxies are monitored; and the regression spectrum of the
\(K=-1\) sector and the intensity correlation of the \(K=0\) sector are
finite-truncation eigenvalue problems whose convergence is checked
numerically, not covered by (v).

The emission spectrum follows from the \(K=-1\) sector by quantum
regression as in Sec.~IV\,E of Paper~I, and the intensity correlation
from the \(K=0\) sector (Sec.~\ref{sec:statistics}); the numerical
solution of the recurrence is described in
Appendices~\ref{app:construction} and \ref{app:normalized}.

\subsection{Observables and photon balance}
\label{sec:obs}

Because \(b_0\) is the only traceful local element and
\(\Tr\rho_n^{(k)}=\delta_{n0}\delta_{k0}\): with \(c_n\) the component of
\(X_n\) along \(Ne_0\), \(\langle\nop\rangle=\nu+(1+\nu)c_1\) and
\(\langle\nop(\nop-1)\rangle=2\nu^2+4\nu(1+\nu)c_1+2(1+\nu)^2c_2\); the
reduced atomic state is \(\rho_N=\sum_{q_\alpha=0}(X_0)_\alpha R_{\mv_\alpha}\);
and for a one-emitter observable \(O\) and a two-emitter product
\(OO'\), with \(o_\mu=\Tr(b_\mu O)\), \(o'_\mu=\Tr(b_\mu O')\),
\begin{align}
&\Big\langle\sum_jO^{(j)}\Big\rangle=N o_0X_0[Ne_0]+\sum_{\mu\neq0}o_\mu X_0[(N-1)e_0+e_\mu],\notag\\
&\Big\langle\sum_{i\neq j}O^{(i)}O'^{(j)}\Big\rangle=N(N-1)\,o_0o'_0\,X_0[Ne_0]\notag\\
&\quad+(N-1)\sum_{\mu\neq0}\big(o_\mu o'_0+o_0o'_\mu\big)\,X_0[(N-1)e_0+e_\mu]\notag\\
&\quad+\sum_{\mu,\nu\neq0}(1+\delta_{\mu\nu})o_\mu o'_\nu X_0[(N-2)e_0+e_\mu+e_\nu],
\label{eq:obsfunctionals}
\end{align}
so that the pair coherence is
\(C_2^{(N)}=\langle\tau_+^{(1)}\tau_-^{(2)}\rangle=X_0[(N-2)e_0+e_++e_-]/[N(N-1)]\).
Joint observables \(\langle a^\dagger O\rangle\) with \(O\) atomic of
charge \(-1\) are \(\sum_{q_\alpha=+1}(X_0)_\alpha\Tr(R_{\mv_\alpha}O)\).

The exact photon balance of Paper~I generalizes with the laser-level
inversion \(W_{ul}=\sum_j(P_u^{(j)}-P_l^{(j)})\) and \(C_{ul}\): with
\(Z=\langle a^\dagger S_-\rangle\),
\begin{align}
A(\langle\nop\rangle-\nu)&=\sum_{i\neq u}\gamma_{u\leftarrow i}\langle N_i\rangle-\Gamma_u\langle N_u\rangle=-g\operatorname{Im}Z,\notag\\
A(\langle\nop\rangle-\nu)&=\mathcal R_{ul}\big[\langle S_+S_-\rangle+\langle\nop W_{ul}\rangle\big],\notag\\
\mathcal R_{ul}&=\frac{g^2(A/2+C_{ul})}{2[(A/2+C_{ul})^2+\Delta^2]},
\label{eq:flux}
\end{align}
where the first line is the stationary balance of the upper-laser-level
population and the second follows from
\(\dot Z=-(A/2+C_{ul}+i\Delta)Z-\tfrac{ig}{2}[\langle S_+S_-\rangle+\langle\nop W_{ul}\rangle]\),
which uses \([\tau_-,\tau_+]=P_l-P_u\) and the eigenvalue of \(\tau_-\)
under the one-emitter Liouvillian.  For \(N=2\),
\(\langle S_+S_-\rangle=\langle N_u\rangle+2\mathcal C_{12}\) and the
singlet--triplet representation \(2\mathcal C_{12}=p_{1,0}-p_{0,0}\) of
Paper~I holds unchanged, the two states being the symmetric and
antisymmetric combinations of \(|ul\rangle,|lu\rangle\): the sign of the
pair coherence is the sign of the population difference between the
bright and the dark pair state within the subspace in which both emitters
occupy laser levels.
\section{Weak pump: pump-quantum filtration}
\label{sec:weakpump}

Paper~I proved, from a filtration of the Liouville space by the total
excitation, that for two-level emitters the joint cumulants
\(K_m=\kappa(n_1,\ldots,n_m)\) of the excitation numbers \(n_j=P_u^{(j)}\)
are \(O(P^m)\) and the connected coherences
\(\Gamma_m=\kappa(\tau_+^{(1)}\tau_-^{(2)},n_3,\ldots,n_m)\) are
\(O(P^{m-1})\) as \(P\to0\) (\(\Gamma_2=C_2\)).  For multilevel schemes the
right grading is not the excitation but the number of pump quanta
absorbed.  Write \(\mathcal L=\mathcal L_0+P\mathcal L_1\) with
\(\mathcal L_1\) the pump dissipators, and suppose that at \(P=0\) the
unique stationary state is \(|0\rangle\langle0|\otimes|g\cdots g\rangle\langle g\cdots g|\)
for one level \(g\).  Call a function \(\chi\) from the levels to the
nonnegative integers with \(\chi(g)=0\) an \emph{admissible charge} if
every pump jump raises \(\chi\) by exactly one, no jump of
\(\mathcal L_0\) raises it, and either (a) \(\chi(u)=\chi(l)+1\), with
\(\Exc_\chi=\nop+\sum_j\chi^{(j)}\), or (b) \(\chi(u)=\chi(l)\), with
\(\Exc_\chi=\sum_j\chi^{(j)}\).  In both cases \(H\) commutes with
\(\Exc_\chi\), the jumps of \(\mathcal L_0\) (including \(a\)) do not
raise it, and the pump jumps raise it by one.  Let
\(V_r=\operatorname{span}\{|i\rangle\langle j|:E_i\le r,\ E_j\le r\}\) in
the eigenbasis of \(\Exc_\chi\), and consider, in any Fock truncation,
the regular expansion \(\rho_{\rm ss}(P)=\sum_rP^r\rho^{(r)}\), which
exists because zero is a simple eigenvalue of the truncated
\(\mathcal L_0\).

\begin{proposition}[Pump-quantum filtration, \(\nu=0\)]
\label{prop:pumpquantum}
For every admissible charge \(\chi\): \(\mathcal L_0V_r\subseteq V_r\),
\(\mathcal L_1V_r\subseteq V_{r+1}\), and \(\rho^{(r)}\in V_r\).
Consequently every operator \(O=\Pi_{\ge k}O\Pi_{\ge k}\) supported on
\(\Exc_\chi\ge k\) has \(\langle O\rangle=O(P^k)\) at the level of the
series, and
\begin{align}
K_m&=O(P^{m\chi_u}),\qquad C_2^{(N)}=O(P^{\chi_u+\chi_l}),\notag\\
\Gamma_m&=O(P^{\chi_u+\chi_l+(m-2)\chi_u}).
\label{eq:pumporders}
\end{align}
The Taylor coefficients \(\rho^{(r)}\) are the same for every Fock cutoff
\(n_{\rm c}>r\) whenever a charge of type (a) is admissible, because
\(V_r\) then contains no Fock level above \(r\); different admissible
charges may be used simultaneously.
\end{proposition}
\begin{proof}
The proof of Lemma~2 of Paper~I uses only that \(H\) commutes with the
grading operator, that every jump of \(\mathcal L_0\) lowers or preserves
it, that every pump jump raises it by at most one, that
\(\rho^{(0)}\in V_0\), and that zero is a simple eigenvalue of
\(\mathcal L_0\) on each finite-dimensional invariant subspace \(V_r\),
which follows from the uniqueness of the stationary state of the
truncated \(\mathcal L_0\).  All hold for \(\Exc_\chi\).  The monomial
\(n_{i_1}\cdots n_{i_k}\) is supported on \(\Exc_\chi\ge k\chi_u\).  The
operator \(\tau_+^{(1)}\tau_-^{(2)}n_{j_1}\cdots n_{j_b}\) maps states
with emitter 1 in \(l\), emitter 2 in \(u\), and \(b\) further emitters in
\(u\) to states with emitter 1 in \(u\), emitter 2 in \(l\), and the same
\(b\) emitters in \(u\); both have \(\Exc_\chi\ge\chi_u+\chi_l+b\chi_u\).
The partition expansion of the cumulants then gives
Eq.~(\ref{eq:pumporders}) term by term, as in Propositions~6 and 7 of
Paper~I.  The last statement holds because \(\Exc_\chi\) of type (a)
bounds \(\nop\) by \(r\) on \(V_r\).
\end{proof}

The admissible charges of the four schemes are listed in
Table~\ref{tab:schemes}.  For the two-level and V schemes the lower laser
level is the ground state, only charges of type (a) give nontrivial
orders, and Eq.~(\ref{eq:pumporders}) reproduces the orders of Paper~I:
\(C_2=O(P)\), \(\Gamma_3=O(P^2)\), \(K_3=O(P^3)\).  For the \(\Lambda\)
scheme and the ladder the lower laser level is itself one pump quantum
above the ground state, a charge of type (b) with \(\chi_u=\chi_l=1\) is
admissible, and the pair coherence is \emph{second} order in the pump,
\(C_2=O(P^2)\), \(\Gamma_3=O(P^3)\), \(K_3=O(P^3)\): two emitters must each
absorb a pump quantum before they can share a coherence in the laser
manifold.  These two schemes also admit type-(a) charges ---
\(\chi_r=\chi_l=0\), \(\chi_u=1\) for \(\Lambda\), and
\(\chi_g=\chi_l=0\), \(\chi_p=\chi_u=1\) for the ladder --- which do not
give the sharpest orders but bound the photon number on \(V_r\), so that
their Taylor coefficients are cutoff independent as well.  (The direct stationary moment bound of Lemma~3 of Paper~I,
which does not rely on the series, needs every level of positive charge
to have a charge-lowering jump; this holds for the \(\Lambda\) scheme
with a type-(a) charge but not otherwise, so the statements above are
statements about the pump series.)

\emph{Leading coefficients from the exact series.}  Throughout, an
``exact'' coefficient means one obtained from the exact finite-system
formulation --- a finite linear system whose solution is the Taylor
coefficient itself, with no closure, fit, or radial truncation error ---
and evaluated numerically in double precision; it is neither a
closed-form expression nor a value with a certified error bound, and
its numerical reliability is assessed by the checks quoted below.  In
the fixed local basis the block-Jacobi matrix is affine in \(P\),
\(\mathcal M(P)=\mathcal M_0+P\mathcal M_1\) (verified to \(3\times10^{-14}\)),
and the finite-system series of Sec.~V\,B of Paper~I applies: the
coefficient vectors obey \(\mathcal M_0X^{(r)}=-\mathcal M_1X^{(r-1)}\)
with vanishing trace component, \(X^{(r)}_n=0\) for \(n>r\), and
observables are power series whose coefficients are
\(\sum_{\mv}X_0^{(r)}[\mv]\Tr(R_{\mv}O)\) with \(P\)-independent traces.
(In the damping basis the matrices are not affine in \(P\), because the
basis itself depends on the rates.)  Table~\ref{tab:weakpump} gives the
leading coefficients for \(N=2,3\) in the bad-cavity regime: all
coefficients below the orders of Eq.~(\ref{eq:pumporders}) vanish to
\(4\times10^{-14}\) and all leading coefficients are nonzero, so the
bounds are attained.  Two features are worth noting.  First, the
leading coefficients of the \(\Lambda\) scheme and the ladder coincide
for every observable in the table, and the ladder coefficients are
independent of the pump-level decay rate (to eight digits for
\(\gamma_p/A=0.3\), \(1\), \(10\), with \(\Lambda\) as the limit
\(\gamma_p\to\infty\)); the two-level model and the V scheme share the
leading photon-number and pair-coherence coefficients, but their leading
higher connected correlations differ and depend on the reservoir decay
of the V scheme (\(\Gamma_3/P^2=0.675\), \(0.715\), \(0.756\), \(0.801\)
for \(\gamma_r/A=0.5\), \(1\), \(2\), \(10\), approaching the two-level
\(0.807\); likewise \(K_3/P^3\)).  We have not derived this pattern in
general, but it is consistent with where the pump acts: in the
\(\Lambda\) scheme and the ladder it acts only on a reservoir that
carries no lasing coherence, whereas in the two-level and V schemes it
empties the lower laser level and thereby acts on the coherences that
the higher connected correlations are built from.  Higher connected
correlations thus retain information about the pump kinetics that the
lower moments do not.  Second, the leading pair-coherence coefficient of the \(\Lambda\) and ladder media
is positive at these parameters, \(C_2=+0.106\,(P/A)^2\), whereas that
of the two-level and V schemes is negative, \(C_2=-1.14\,(P/A)\) at
\(N=2\): Paper~I traced the negative two-level coherence to the trapping
of population in the dark pair singlet, which only local pump and decay
can enter or leave; a lower laser level emptied by recycling depletes
the singlet, which requires one emitter in \(l\).  For the test posed in
Sec.~\ref{sec:intro} this is a qualitative failure: since
\(\langle\nop\rangle\) is linear in \(P\) at weak pump for every scheme,
fitting the two-level pump rate to the photon number gives
\(C_2\simeq-1.67\langle\nop\rangle\) for two two-level emitters, against
\(C_2\simeq+0.079\langle\nop\rangle^2\) for two \(\Lambda\) or ladder
emitters at these parameters --- wrong in sign and in order.

\begin{table*}[tbp]
\caption{Leading weak-pump coefficients from the exact finite-system
series in the bad-cavity regime (\(g=1.1A\), \(\gamma=0.3A\),
\(\gamma_r=\gamma_p=A\), \(\nu=0\)), with \(P\) in units of \(A\):
\(\langle\nop\rangle=n_1P+\ldots\), \(C_2=c_kP^k+\ldots\),
\(\Gamma_3=\gamma_kP^k+\ldots\), \(K_3=\kappa_3P^3+\ldots\).  The largest
coefficient below the leading order is \(4\times10^{-14}\) (two-level and
V, \(N=3\)) and below \(10^{-14}\) otherwise; the affine residual of the
fixed-basis matrix and the violation of \(X^{(r)}_n=0\) for \(n>r\) are
below \(3\times10^{-13}\) in every case.}
\label{tab:weakpump}
\begin{ruledtabular}
\begin{tabular}{lcrrrr}
scheme & \(N\) & \(n_1\) & \(C_2\) & \(\Gamma_3\) & \(K_3\)\\
\hline
two-level & 2 & 0.68439 & \(-1.14065\,P\) & & \\
V & 2 & 0.68439 & \(-1.14065\,P\) & & \\
\(\Lambda\) & 2 & 1.16262 & \(+0.10618\,P^2\) & & \\
ladder & 2 & 1.16262 & \(+0.10618\,P^2\) & & \\
two-level & 3 & 0.71051 & \(-0.78946\,P\) & \(+0.80742\,P^2\) & \(+5.20465\,P^3\)\\
V & 3 & 0.71051 & \(-0.78946\,P\) & \(+0.71481\,P^2\) & \(+6.08860\,P^3\)\\
\(\Lambda\) & 3 & 1.74393 & \(+0.10618\,P^2\) & \(+0.01939\,P^3\) & \(-0.00929\,P^3\)\\
ladder & 3 & 1.74393 & \(+0.10618\,P^2\) & \(+0.01939\,P^3\) & \(-0.00929\,P^3\)\\
\end{tabular}
\end{ruledtabular}
\end{table*}

\begin{figure*}[tbp]
\includegraphics[width=\textwidth]{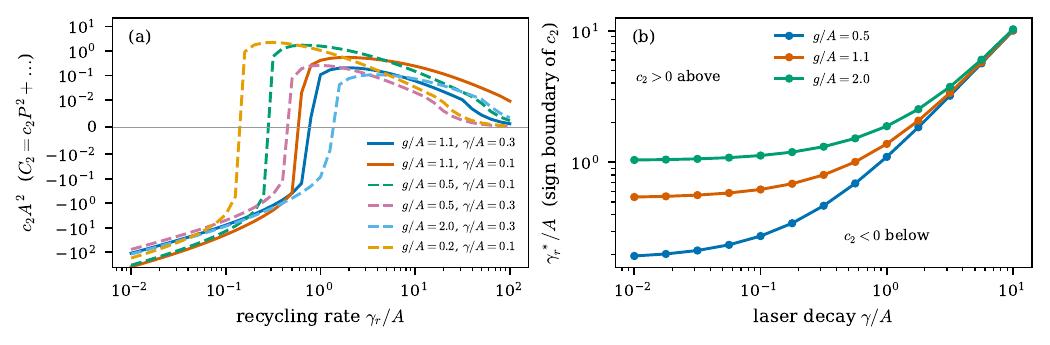}
\caption{Leading pair-coherence coefficient \(c_2\) of two \(\Lambda\)
emitters, \(C_2=c_2P^2+O(P^3)\), from the exact finite-system series
(evaluated numerically) at \(\Delta=0\),
\(\nu=0\): (a) versus recycling rate for several couplings and
laser-transition decays (symmetric-log scale); (b) the sign boundary
\(\gamma_r^*\) versus \(\gamma\) for three couplings; \(c_2>0\) above the
curves.}
\label{fig:signmap}
\end{figure*}

\emph{Sign boundary.}  The filtration fixes the order of the pair
coherence, not its sign, and the sign depends on the recycling rate.
Figure~\ref{fig:signmap} maps the leading coefficient \(c_2\) of two
\(\Lambda\) emitters, \(C_2=c_2P^2+O(P^3)\), as a function of
\(\gamma_r\), \(g\), and \(\gamma\) (the ladder with
\(\gamma_p=\gamma_r\) has the same \(c_2\)).  At slow recycling the
lower laser level is populated, the singlet traps population as for
two-level emitters, and \(c_2<0\); above a boundary \(\gamma_r^*\) the
coherence is positive --- a bright-triplet excess, i.e.\ constructive
collective interference in the sense of Eq.~(\ref{eq:flux}), not a
statement about superradiant scaling of the output.  The boundary is
\(\gamma_r^*\approx\gamma\) when the laser-transition decay dominates
(\(\gamma_r^*/A=10.0\), \(10.1\), \(10.3\) at \(\gamma=10A\) for
\(g/A=0.5,1.1,2\)) and saturates at a coupling-dependent value when
\(\gamma\to0\) (\(\gamma_r^*/A=0.19\), \(0.54\), \(1.04\)).  At the
good-cavity parameters of Sec.~\ref{sec:strongpump} (\(g=5A\),
\(\gamma=A\)) the weak-pump boundary lies at \(\gamma_r^*=3.547A\),
between the two ladders of Fig.~\ref{fig:pumpscan}: the fast one
(\(\gamma_r=10A\)) has a positive and the slow one (\(\gamma_r=3A\)) a
negative pair coherence over the whole pump range.  The finite-pump sign
change lies near this value and shifts with pump and emitter number,
by at most \(9\%\) for the cases checked (\(N=2,3\) up to \(P=16A\),
where Fig.~\ref{fig:recycling} is computed; Appendix~\ref{app:validation}, item (i)).

\section{Strong pump: self-quenching versus saturation}
\label{sec:strongpump}

\subsection{Which schemes quench}

The flux identity (\ref{eq:flux}) gives a sufficient condition for
strong-pump quenching.

\begin{proposition}[Pump-induced quenching, \(\nu=0\)]
\label{prop:quench}
Let \(\nu=0\) and keep \(A>0\), \(g\), \(\Delta\), and all pump-independent
rates fixed.  If the pump empties a laser level --- \(P\) enters
\(\Gamma_u\) or \(\Gamma_l\), so that \(C_{ul}\to\infty\) as \(P\to\infty\)
--- then
\begin{equation}
\langle\nop\rangle\le\frac{N^2\mathcal R_{ul}}{A-N\mathcal R_{ul}},\qquad
\mathcal R_{ul}\le\frac{g^2}{A+2C_{ul}},
\label{eq:quenchbound}
\end{equation}
once \(N\mathcal R_{ul}<A\), so that \(\langle\nop\rangle\to0\) as
\(P\to\infty\), in every Fock truncation and, under (U), for
\(\rho_{\rm ss}\).
\end{proposition}
\begin{proof}
In a truncation to Fock levels \(0,\ldots,M-1\), with \(a\) the compressed
annihilation operator, \([\nop,a^\dagger]=a^\dagger\) and the cavity-loss
terms of the equations for \(\langle\nop\rangle\) and \(Z\) are unchanged,
so the first line of Eq.~(\ref{eq:flux}) holds as it stands; the only
change is \([a,a^\dagger]=\openone-M\Pi_{M-1}\), with \(\Pi_{M-1}\) the
projector on the top Fock level, which adds \(-M\langle\Pi_{M-1}S_+S_-\rangle\)
to the bracket in the second line.  Since \(\Pi_{M-1}\) and \(S_+S_-\) are
commuting positive operators, this term is nonpositive, and in every
truncation as well as for the untruncated equation
\(A\langle\nop\rangle\le\mathcal R_{ul}[\langle S_+S_-\rangle+\langle\nop W_{ul}\rangle]
\le\mathcal R_{ul}[N^2+N\langle\nop\rangle]\), using \(\|S_+S_-\|\le N^2\) and
\(W_{ul}\le N\) commuting with \(\nop\ge0\).  The bound on
\(\mathcal R_{ul}\) follows from Eq.~(\ref{eq:flux}) by dropping
\(\Delta^2\).  For \(\rho_{\rm ss}\) the identities hold because its
moments are finite (Proposition~\ref{prop:carry}(v)).
\end{proof}

For a thermal cavity, \(\nu>0\), the analogous two-sided bounds on
\(\langle\nop\rangle-\nu\) give \(\langle\nop\rangle\to\nu\) for the
untruncated equation, but only for a stationary state with finite second
moments, which hypothesis (U) does not provide; in a finite truncation the limit is instead the truncated
thermal mean \(\nu-Mr^M/(1-r^M)\), \(r=\nu/(1+\nu)\), and the flux identity
acquires boundary terms from the thermal pump.  We do not pursue this
case.

Pumping out of a laser level is a sufficient mechanism for quenching, not
a necessary one.  A pump can also starve the laser transition without
broadening it, by diverting population into a shelving cycle: with
levels \(a,b,u,l\), pump \(a\to b\) at rate \(P\), and fixed rates
\(b\to a\), \(a\to u\), \(u\to l\), \(l\to a\) all equal to \(A\), the rate
graph is irreducible and \(C_{ul}=A\) is pump independent, yet the
single-emitter photon number at \(g=1.1A\) falls from \(0.0675\) at
\(P=A\) to \(2.5\times10^{-4}\) at \(P=1000A\), roughly as \(1/P\), because the feeding current \(a\to u\) vanishes as
population accumulates in \(b\).  What saturation requires is that the
pump does not reduce the injection current into the upper laser level;
the \(\Lambda\) and ladder cycles studied here satisfy this.

For the \(\Lambda\) scheme and the ladder, \(C_{ul}\) is pump
independent and the limit is a different model: as \(P\to\infty\) the
pumped level is emptied instantly, so the \(\Lambda\) scheme becomes the
two-level model with pump rate \(\gamma_r\) (the recycling jump followed
by instantaneous re-excitation) and the ladder becomes the V scheme with
pump \(\gamma_r\) and reservoir decay \(\gamma_p\).  Numerically, at
\(N=2\) in the good-cavity regime, the \(\Lambda\) laser with
\(\gamma_r=10A\) reaches the photon number \(6.678\) of its limiting
two-level model to \(10^{-3}\) at \(P=6400A\), and the ladder its V
limit \(4.342\) likewise, while the two-level and V photon numbers fall
as \(1/P\) (Appendix~\ref{app:validation}, item (i)).  The approach to the finite limit is monotonic for the
cycles studied here; for other graphs neither monotonicity nor a nonzero
limit is implied.

\subsection{Exact comparison through the lasing crossover}

\begin{figure*}[tbp]
\includegraphics[width=\textwidth]{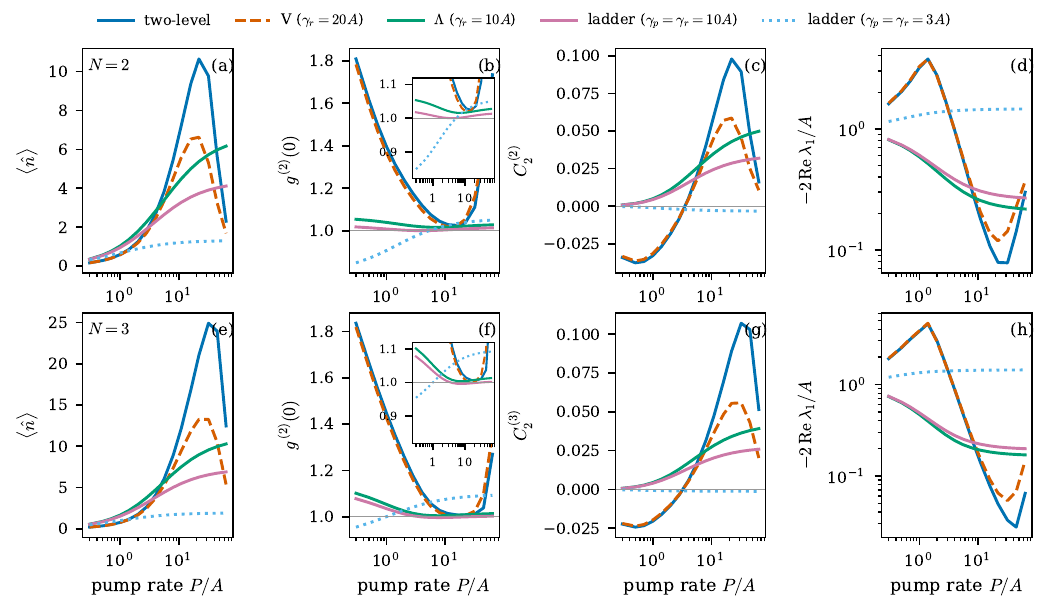}
\caption{Pump-scheme dependence of two- and three-emitter lasing in the
good-cavity regime \(g=5A\), \(\gamma=A\), \(\Delta=0\), \(\nu=0\), versus
pump rate: (a,e) mean photon number, (b,f) \(g^{(2)}(0)\) (insets: the
region near unity), (c,g) pair coherence \(C_2^{(N)}\), (d,h) width
\(-2\operatorname{Re}\lambda_1\) of the dominant spectral pole; top row
\(N=2\), bottom row \(N=3\).  Two-level model; V scheme with
\(\gamma_r=20A\); \(\Lambda\) scheme with \(\gamma_r=10A\); ladder with
\(\gamma_p=\gamma_r=10A\) and with \(\gamma_p=\gamma_r=3A\).  Thin grey
lines mark \(g^{(2)}(0)=1\) and \(C_2=0\).  All points are converged in
\(n_{\max}\) (\(n_{\max}=70\); the largest-photon-number point of the
scan, \(N=3\), two-level, \(P/A=32\), changes by \(4\times10^{-10}\) between
\(n_{\max}=70\) and \(120\) and agrees with a brute-force Fock-space
solution to eight digits, Appendix~\ref{app:validation}).}
\label{fig:pumpscan}
\end{figure*}

Figure~\ref{fig:pumpscan} follows \(N=2\) and \(N=3\) emitters through
the lasing crossover for the four schemes at fixed laser-transition
parameters and increasing pump rate.  In the two-level model and the V
scheme the photon number passes through a maximum (largest sampled
values in Fig.~\ref{fig:pumpscan}: \(\langle\nop\rangle=10.65\) at \(P/A=22\)
for \(N=2\) and \(24.9\) at \(P/A=32\) for \(N=3\), \(6.6\) and \(13.3\)
in the V scheme; maximizing over \(P\) gives \(10.77\) at
\(P/A=24.5\) and \(25.36\) at \(P/A=36.7\) for the two-level model and
\(6.70\) and \(13.58\) for the V scheme) and then
collapses, to \(2.3\) and \(12.5\) (two-level) and \(1.7\) and \(5.4\) (V)
at \(P/A=64\), while \(g^{(2)}(0)\) rises back towards thermal values and
the pair coherence, positive and growing through the lasing regime,
falls back towards zero.  This is the self-quenching of incoherently
pumped one-atom lasers \cite{MuSavage1992}, here with its collective
signature, and Proposition~\ref{prop:quench} shows it is unavoidable for
these two schemes at zero cavity temperature.  In the \(\Lambda\) scheme and the ladder nothing
quenches: the photon number, the inversion, and the linewidth saturate
monotonically towards their limiting models, \(g^{(2)}(0)\) passes
through a shallow minimum near \(P/A\approx8\) and stays within
\(1.0\)--\(1.1\) for the \(\Lambda\) scheme and the fast ladder (minimum
\(0.996\)), and the pair coherence saturates, positive for
\(\gamma_r=10A\) and negative over the whole range for the slow ladder
(Sec.~\ref{sec:weakpump}).  At weak pump the \(\Lambda\) and ladder
lasers are inverted from the start, because the lower laser level is
emptied by recycling rather than filled by the pump: at \(P/A=0.5\)
their photon numbers are roughly two to two and a half times those of
the two-level model (\(0.57\) and \(0.54\) against \(0.28\) at \(N=2\);
\(0.89\) and \(0.85\) against \(0.34\) at \(N=3\)) and their equal-time
statistics are already near-Poissonian (\(g^{(2)}(0)=1.05\), \(1.01\)) where the
two-level model is still thermal-like (\(1.6\)).

\begin{figure*}[tbp]
\includegraphics[width=\textwidth]{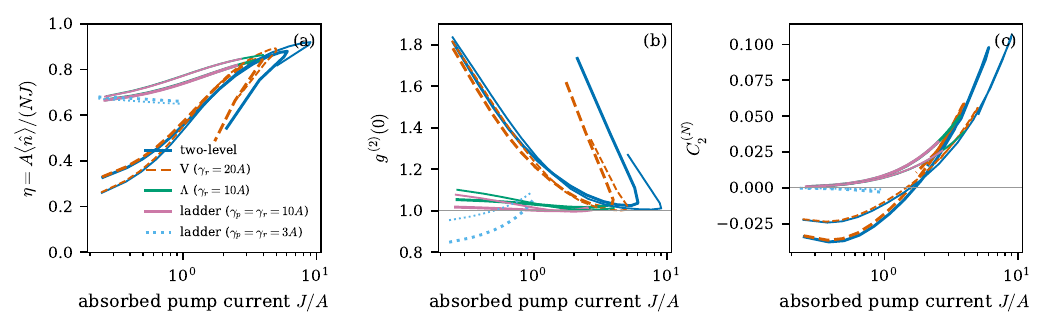}
\caption{The data of Fig.~\ref{fig:pumpscan} against the absorbed pump
current per emitter \(J=Pp_{\rm origin}\): (a) conversion fraction
\(\eta=A\langle\nop\rangle/(NJ)\), (b) \(g^{(2)}(0)\), (c) pair
coherence; thick lines \(N=2\), thin lines \(N=3\).  The two-level and V
curves turn around because \(J\) decreases again at strong pump.}
\label{fig:current}
\end{figure*}

Comparisons at fixed bare pump rate are not comparisons at fixed
excitation: the pump absorbed per emitter is \(J=Pp_{\rm origin}\), with
\(p_{\rm origin}\) the population of the level the pump empties, and the
schemes differ in \(p_{\rm origin}\).  Figure~\ref{fig:current} replots
the same data against \(J\), showing the conversion fraction
\(\eta=A\langle\nop\rangle/(NJ)\), the share of the absorbed excitation
current that leaves through the cavity --- an excitation-current
conversion fraction, not a pump-power or wall-plug efficiency, since
neither the pump photon energy nor the losses of the pump source enter
it: by the first line of
Eq.~(\ref{eq:flux}), \(A\langle\nop\rangle=NJ-\gamma\langle N_u\rangle\) in
stationarity, and \(1-\eta\) is the share lost by spontaneous decay of the
laser transition.  Matching the absorbed current reduces some of the
differences seen at fixed pump rate, but appreciable ones remain because
the spontaneous loss depends on the level scheme.  At \(J=A\) the photon
numbers are \(1.112\), \(1.142\), \(1.481\), \(1.478\) (two-level, V,
\(\Lambda\), fast ladder) for \(N=2\) and \(1.604\), \(1.655\), \(2.321\),
\(2.319\) for \(N=3\), i.e.\ \(\eta=0.56\), \(0.57\), \(0.74\), \(0.74\)
and \(0.53\), \(0.55\), \(0.77\), \(0.77\): the conversion fraction of
the \(\Lambda\) and ladder lasers exceeds that of the two-level laser
by \(33\%\) and \(45\%\) at \(N=2\) and \(3\),
because their lower laser level is kept empty by recycling and the
cavity competes with \(\gamma\) from an inverted transition.  Quenching
appears in Fig.~\ref{fig:current} as the absorbed current of the
two-level and V schemes turning back once the lower laser level is
emptied: it reaches a maximum of \(J/A=6.1\) and \(9.2\) for the
two-level laser and \(4.0\) and \(5.1\) for the V laser (\(N=2,3\)), while
the \(\Lambda\) and ladder currents keep growing to the recycling
bottleneck.  Figure~\ref{fig:pumpscan} is therefore a parameter-matched
illustration, not an efficiency ranking.

\subsection{Linewidth and recycling}

\begin{figure}[tbp]
\includegraphics[width=\linewidth]{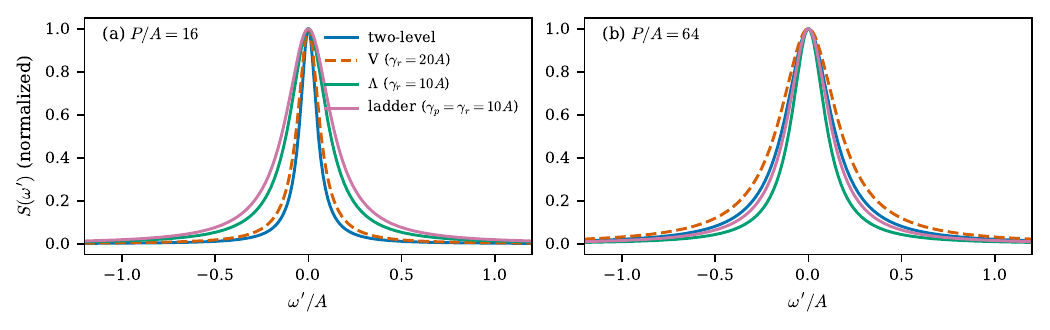}
\caption{Normalized emission spectra \(S(\omega')\) of the \(N=2\) lasers
of Fig.~\ref{fig:pumpscan} at (a) \(P/A=16\) and (b) \(P/A=64\), versus
detuning from the cavity frequency, from the pole decomposition of the
\(K=-1\) continued fraction (\(n_{\max}=60\)).}
\label{fig:spectra}
\end{figure}

The dominant-pole width (Fig.~\ref{fig:pumpscan}d,h; the selection of
the dominant pole is insensitive to the residue threshold,
Appendix~\ref{app:validation}) is, over the lasing range, well described
by its inverse proportionality to the photon number, so that at equal
\(P\) the scheme with the larger \(\langle\nop\rangle\) has the narrower
line: at \(P/A=16\) the widths are \(0.108A\), \(0.138A\), \(0.241A\),
\(0.292A\) for the two-level, V, \(\Lambda\), and ladder
(\(\gamma_p=\gamma_r=10A\)) schemes at \(N=2\), and \(0.077A\), \(0.091A\),
\(0.182A\), \(0.212A\) at \(N=3\).  The schemes that do not quench do not
have narrower lines at fixed pump in this regime; their advantage is
that the line does not rebroaden at strong pump, where the two-level
and V linewidths pass through a minimum (\(0.078A\) and \(0.119A\) at
\(N=2\), \(0.027A\) and \(0.053A\) at \(N=3\)) and then grow again
through the quenching transition, by a factor of three to four at
\(N=2\) and two to three at \(N=3\) by \(P/A=64\).
Figure~\ref{fig:spectra} shows the spectral functions at \(P/A=16\) and
\(64\): at \(P/A=16\) the two-level line is the narrowest (full width at
half maximum \(0.100A\), against \(0.133A\), \(0.233A\), \(0.283A\)),
while at \(P/A=64\) the order is reversed, the quenched two-level and V
lines having broadened to \(0.29A\) and \(0.37A\) while the \(\Lambda\)
and ladder lines, \(0.22A\) and \(0.27A\), are essentially unchanged.
The FWHM agrees with the dominant-pole width to \(3\)--\(7\%\) at
\(P/A=16\) and \(0.5\)--\(3\%\) at \(P/A=64\).

\begin{figure*}[tbp]
\includegraphics[width=\textwidth]{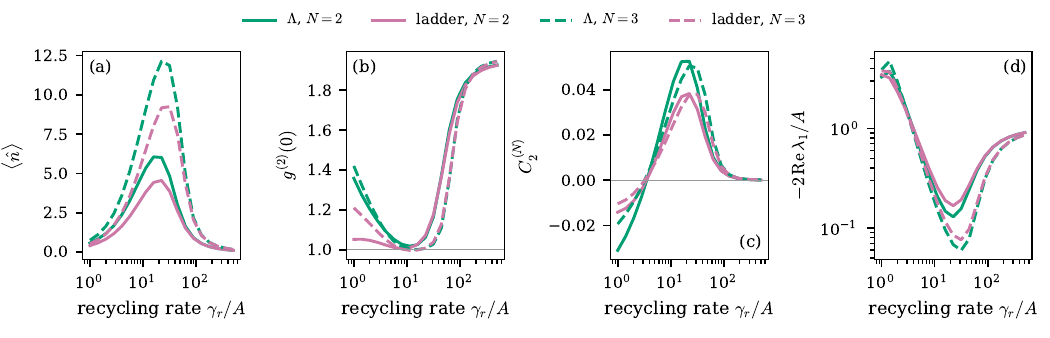}
\caption{Dependence on the recycling rate at fixed pump \(P=16A\) in
the good-cavity regime, for the \(\Lambda\) scheme and the ladder
(\(\gamma_p=\gamma_r\)), \(N=2\) (solid) and \(N=3\) (dashed): (a) photon
number, (b) \(g^{(2)}(0)\), (c) pair coherence, (d) dominant-pole width.}
\label{fig:recycling}
\end{figure*}

For the \(\Lambda\) scheme and the ladder the recycling rate plays the
role that \(P\) plays for the V scheme in \(C_{ul}\): it must be fast
enough to empty the lower laser level but not so fast as to dephase the
coherence.  Figure~\ref{fig:recycling} scans it at fixed \(P=16A\).
Photon number and linewidth are both optimized at recycling rates well
above the laser-transition decay and comparable to or above the pump,
\(\gamma_r/A\approx16\)--\(32\), but not at the same value: the \(N=2\)
\(\Lambda\) laser reaches \(\langle\nop\rangle=6.05\) at \(\gamma_r/A=16\)
and its narrowest line, \(0.129A\), at \(\gamma_r/A=22\); the \(N=3\)
\(\Lambda\) laser \(12.1\) at \(\gamma_r/A=22\) and \(0.060A\) at
\(\gamma_r/A=32\); the ladder (\(\gamma_p=\gamma_r\)) \(4.55\) and
\(0.167A\) at \(\gamma_r/A=22\) for \(N=2\), and \(9.24\) and \(0.076A\) at
\(\gamma_r/A=32\) for \(N=3\).  On either side the laser degrades for
opposite reasons: at slow recycling the lower laser level fills, the
inversion is lost, the statistics become bunched (\(g^{(2)}(0)\) up to
\(1.4\) at \(N=3\)), and the pair coherence turns negative below
\(\gamma_r/A=3.5\) (\(N=2\)) and \(3.25\) (\(N=3\)), near the weak-pump
boundary of Fig.~\ref{fig:signmap} (Sec.~\ref{sec:weakpump}); at fast
recycling \(C_{ul}=(\gamma+\gamma_r)/2\)
dephases the coherence, the photon number collapses, and the field
returns to thermal statistics, \(g^{(2)}(0)\to2\).  The exact solution
thus allows the recycling rate of a given few-emitter device to be
optimized for photon number or for linewidth.

\section{Photon statistics and the pump cycle}
\label{sec:statistics}

A value \(g^{(2)}(0)<1\) establishes a sub-Poissonian intracavity photon
distribution, \(F_n=\operatorname{Var}(\nop)/\langle\nop\rangle=1+\langle\nop\rangle[g^{(2)}(0)-1]<1\).
It does not by itself establish reduced noise in the detected output,
which depends on the time integral of \(g^{(2)}(\tau)-1\): for ideal
detection of the cavity-loss channel, the Fano factor of the photon
count in a long interval is
\begin{equation}
F_{\rm count}=1+2A\langle\nop\rangle\int_0^\infty\big[g^{(2)}(\tau)-1\big]d\tau .
\label{eq:Fcount}
\end{equation}
With a finite overall detection efficiency \(\eta\) (collection and
detector), the counts are a binomial thinning of the emitted photons and
the measured Fano factor is \(1+\eta(F_{\rm count}-1)\): the suppression
is reduced in proportion to \(\eta\) but keeps its sign.
Both are available exactly.  The intensity correlation
\(\langle a^\dagger(0)a^\dagger(\tau)a(\tau)a(0)\rangle=\Tr[\nop\,e^{\mathcal L\tau}(a\rho_{\rm ss}a^\dagger)]\)
evolves the permutation-invariant, charge-zero operator
\(a\rho_{\rm ss}a^\dagger\) under the \(K=0\) block-Jacobi operator; its
coefficients follow from those of \(\rho_{\rm ss}\) by two field
identities, the functional \(\Tr(\nop\,\cdot\,)\) picks the \(n=1\),
\(Ne_0\) component, and the pole decomposition of the truncated \(K=0\)
operator gives \(g^{(2)}(\tau)=\sum_jw_je^{\lambda_j\tau}/\langle\nop\rangle^2\),
with the \(\lambda=0\) term equal to \(\langle\nop\rangle^2\) and
\(F_{\rm count}=1+(2A/\langle\nop\rangle)\sum_{\lambda_j\neq0}w_j/(-\lambda_j)\).
Because the residues of a non-normal operator carry rounding errors, we
evaluate the integral in Eq.~(\ref{eq:Fcount}) instead from the Poisson
equation \(\mathcal M Z=-(Y-\langle\nop\rangle X_{\rm ss})\), \(\Tr Z=0\),
for the truncated \(K=0\) block-Jacobi matrix \(\mathcal M\), the
coefficient vector \(Y\) of \(a\rho_{\rm ss}a^\dagger\), and the
stationary vector \(X_{\rm ss}\); this is independent of the radial
cutoff to nine digits, whereas the pole sum scatters at the \(10^{-6}\)
level.  An independent calculation in the full Fock\(\otimes\)emitter
space (no symmetry reduction, no damping basis, Poisson equation for the
Liouvillian) reproduces \(F_{\rm count}\) to \(10^{-12}\) or better for
one, two, and three emitters (Appendix~\ref{app:validation}).

Figure~\ref{fig:pumpscan}b,f already shows that the ladder with slow
steps, \(\gamma_p=\gamma_r=3A\), has \(g^{(2)}(0)<1\) over the
weak-to-intermediate pump range (\(P\lesssim4A\) at \(N=2\),
\(P\lesssim0.7A\) at \(N=3\), where \(\langle\nop\rangle\lesssim1\)),
\(0.849\) at \(P=0.3A\) for \(N=2\) and \(0.955\) for \(N=3\), and that
the two-level, V, and \(\Lambda\) schemes of that figure stay above one
throughout.  Such a comparison at fixed pump rate does not isolate the
mechanism, because the schemes have different photon numbers and
different absorbed currents.  Table~\ref{tab:pumpstep} therefore
makes the test of Sec.~\ref{sec:intro} explicit.  Held fixed in every
row are the cavity (\(A\), \(\Delta=0\), \(\nu=0\)), the coupling
\(g=5A\), the laser-transition decay \(\gamma=A\), and \(N\); the only
fitted parameter is the pump rate \(P\), chosen so that every scheme
gives the same \(\langle\nop\rangle\); the observables compared after
the fit are \(g^{(2)}(0)\), \(F_n\), and \(F_{\rm count}\).  It
compares the ladder with a fixed recycling
rate \(\gamma_r=3A\) and a pump-level decay \(\gamma_p\) that is varied
from slow to fast --- \(\gamma_p\to\infty\) is the \(\Lambda\) scheme
with the ground level as reservoir --- against the V scheme with the
same \(\gamma_r\), the V scheme with fast reservoir decay, and the
two-level model, for \(N=1,2,3\) and \(\langle\nop\rangle=0.3\) and \(1\).

\begin{table*}[tbp]
\caption{Photon statistics at matched photon number in the good-cavity
regime (\(g=5A\), \(\gamma=A\)): pump rate \(P\) at which
\(\langle\nop\rangle\) equals the target --- the smallest such \(P\), i.e.\
the rising branch below the photon-number maximum of the two-level and V
schemes, found by bisection after a coarse scan --- \(g^{(2)}(0)\), intracavity Fano
factor \(F_n\), and long-time output-counting Fano factor \(F_{\rm count}\),
Eq.~(\ref{eq:Fcount}).  Ladder: \(\gamma_r=3A\) and the pump-level
decay \(\gamma_p\) indicated; \(\Lambda\): \(\gamma_r=3A\) (the
\(\gamma_p\to\infty\) limit of this ladder); V: \(\gamma_r=3A\) or
\(20A\).  At \(\langle\nop\rangle=1\) the intracavity Fano factor
\(F_n=1+\langle\nop\rangle[g^{(2)}(0)-1]\) equals \(g^{(2)}(0)\) and is not
listed.  ``---'': the target photon number is not reached at any pump.}
\label{tab:pumpstep}
\begin{ruledtabular}
\setlength{\tabcolsep}{5.5pt}
\begin{tabular}{lrrrrcrrr}
 & \multicolumn{4}{c}{\(\langle\nop\rangle=0.3\)} & \hspace{1.2em} & \multicolumn{3}{c}{\(\langle\nop\rangle=1\)}\\
scheme, \(N\) & \(P/A\) & \(g^{(2)}(0)\) & \(F_n\) & \(F_{\rm count}\) & & \(P/A\) & \(g^{(2)}(0)\) & \(F_{\rm count}\)\\
\hline
ladder, $\gamma_p=A$, 1 & 1.52 & 0.549 & 0.865 & 0.547 &  & --- & --- & ---\\
ladder, $\gamma_p=3A$, 1 & 0.76 & 0.597 & 0.879 & 0.605 &  & --- & --- & ---\\
ladder, $\gamma_p=10A$, 1 & 0.64 & 0.674 & 0.902 & 0.684 &  & --- & --- & ---\\
ladder, $\gamma_p=100A$, 1 & 0.61 & 0.719 & 0.916 & 0.723 &  & 29.84 & 1.020 & 0.921\\
$\Lambda$ ($\gamma_p\to\infty$), 1 & 0.60 & 0.725 & 0.918 & 0.727 &  & 22.98 & 1.021 & 0.922\\
V, $\gamma_r=3A$, 1 & 0.86 & 0.816 & 0.945 & 0.804 &  & --- & --- & ---\\
V, $\gamma_r=20A$, 1 & 0.73 & 0.987 & 0.996 & 0.907 &  & 3.03 & 1.015 & 0.912\\
two-level, 1 & 0.72 & 1.033 & 1.010 & 0.927 &  & 2.79 & 1.064 & 0.995\\
\hline
ladder, $\gamma_p=A$, 2 & 0.35 & 0.783 & 0.935 & 0.663 &  & --- & --- & ---\\
ladder, $\gamma_p=3A$, 2 & 0.28 & 0.847 & 0.954 & 0.775 &  & 3.05 & 0.974 & 0.740\\
ladder, $\gamma_p=10A$, 2 & 0.26 & 0.893 & 0.968 & 0.829 &  & 1.78 & 0.990 & 0.779\\
ladder, $\gamma_p=100A$, 2 & 0.26 & 0.917 & 0.975 & 0.852 &  & 1.53 & 1.007 & 0.815\\
$\Lambda$ ($\gamma_p\to\infty$), 2 & 0.26 & 0.920 & 0.976 & 0.855 &  & 1.51 & 1.010 & 0.819\\
V, $\gamma_r=3A$, 2 & 0.57 & 1.394 & 1.118 & 1.044 &  & 2.12 & 1.099 & 0.971\\
V, $\gamma_r=20A$, 2 & 0.53 & 1.562 & 1.169 & 1.124 &  & 1.69 & 1.242 & 1.200\\
two-level, 2 & 0.53 & 1.595 & 1.178 & 1.138 &  & 1.63 & 1.279 & 1.248\\
\hline
ladder, $\gamma_p=A$, 3 & 0.20 & 0.893 & 0.968 & 0.754 &  & 3.28 & 1.003 & 0.759\\
ladder, $\gamma_p=3A$, 3 & 0.17 & 0.940 & 0.982 & 0.847 &  & 1.03 & 1.001 & 0.752\\
ladder, $\gamma_p=10A$, 3 & 0.17 & 0.972 & 0.992 & 0.887 &  & 0.83 & 1.024 & 0.818\\
ladder, $\gamma_p=100A$, 3 & 0.16 & 0.989 & 0.997 & 0.903 &  & 0.77 & 1.038 & 0.853\\
$\Lambda$ ($\gamma_p\to\infty$), 3 & 0.16 & 0.991 & 0.997 & 0.905 &  & 0.76 & 1.040 & 0.858\\
V, $\gamma_r=3A$, 3 & 0.47 & 1.579 & 1.174 & 1.107 &  & 1.49 & 1.237 & 1.124\\
V, $\gamma_r=20A$, 3 & 0.45 & 1.683 & 1.205 & 1.149 &  & 1.29 & 1.353 & 1.277\\
two-level, 3 & 0.44 & 1.704 & 1.211 & 1.157 &  & 1.26 & 1.381 & 1.308\\
\end{tabular}
\end{ruledtabular}
\end{table*}

\begin{figure*}[tbp]
\includegraphics[width=\textwidth]{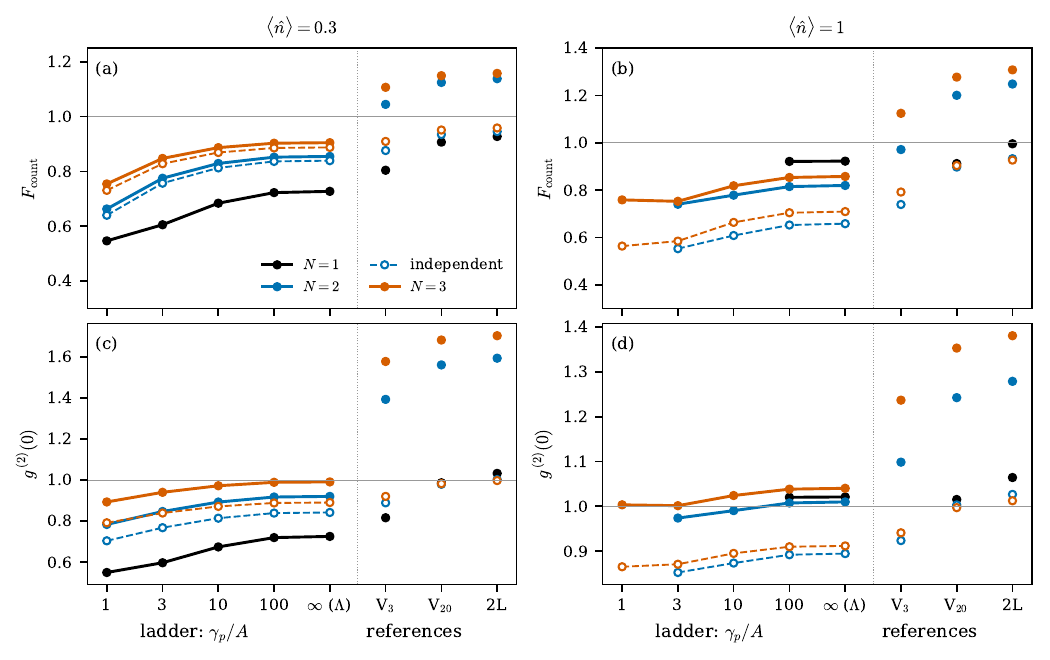}
\caption{The matched-photon-number comparison of Table~\ref{tab:pumpstep}
(\(g=5A\), \(\gamma=A\)): (a,b) long-time output-counting Fano factor and
(c,d) \(g^{(2)}(0)\) at \(\langle\nop\rangle=0.3\) (left) and \(1\) (right),
for the ladder with \(\gamma_r=3A\) and pump-level decay \(\gamma_p\)
(connected points; \(\gamma_p\to\infty\) is the \(\Lambda\) scheme) and
for the reference schemes (V with \(\gamma_r=3A\), V with \(\gamma_r=20A\),
two-level), for \(N=1,2,3\) emitters.  Open symbols (dashed): \(N\)
independent single-emitter lasers of the same scheme, each in its own
cavity, at the same total photon number --- the counting Fano factor of
their combined output is that of one of them, and
\(g^{(2)}_{\rm tot}(0)=[g^{(2)}_1(0)+N-1]/N\).  Missing points: the target
photon number is not reached.  Thin lines mark unity.}
\label{fig:pumpstep}
\end{figure*}

The table separates the two effects.  At matched
\(\langle\nop\rangle=0.3\), every statistic decreases monotonically as the
pump step is slowed (at \(\langle\nop\rangle=1\) the same holds except for
the last step \(\gamma_p=3A\to A\) at \(N=3\), where the three statistics
rise by \(0.002\)--\(0.007\)): for
\(N=2\) at \(\langle\nop\rangle=0.3\), \(F_{\rm count}\) falls from
\(0.855\) (\(\Lambda\), no pump-level bottleneck) through \(0.852\),
\(0.829\), \(0.775\) to \(0.663\) as \(\gamma_p\) goes from \(100A\) to \(A\),
and for \(N=3\) from \(0.905\) to \(0.754\); at \(\langle\nop\rangle=1\), where
\(g^{(2)}(0)\) is within a few percent of one, the counting Fano factor
is still \(0.74\)--\(0.82\) for \(N=2\) and \(0.75\)--\(0.86\) for \(N=3\),
so the output counting noise is reduced well into the regime where the
equal-time statistics are near-Poissonian.  The reference schemes order the
same way: the V scheme with a slow reservoir decay (\(\gamma_r=3A\)), a
two-step cycle, is quieter than the V scheme with \(\gamma_r=20A\),
which is in turn quieter than the two-level model, whose excitation of
the upper laser level is a single Poissonian jump --- but for \(N\ge2\)
the two-level model and the fast V scheme stay above
\(F_{\rm count}=1\) in every row, the slow V scheme dips below it only
at \(N=2\), \(\langle\nop\rangle=1\) (\(0.97\)), whereas every ladder and
the \(\Lambda\) scheme is below one in every row.  (A single two-level
emitter has \(F_{\rm count}=0.93\) at \(\langle\nop\rangle=0.3\) although
its \(g^{(2)}(0)=1.03\): the counting statistics integrates
\(g^{(2)}(\tau)-1\) over the emitter's recovery time.)  The
picture that emerges is the one of dynamic pump-noise suppression
\cite{Ritsch1991,RitschZoller1992,Briegel1996}, the multilevel counterpart
of regular pumping \cite{Yamamoto1986,Haake1989}: rate-limited steps in the excitation
cycle make the arrival of excitations in the upper laser level more
regular than a Poisson process, and a long-lived cavity field inherits
the regularity.  Here it is obtained exactly for a few emitters from the
full level kinetics, rather than from an effective pump operator; the
treatment of Ref.~\cite{Briegel1996}, which derives a non-Markovian,
photon-number-dependent pump operator for a single multilevel atom, is
the one-atom counterpart.  Figure~\ref{fig:pumpstep} shows the content
of Table~\ref{tab:pumpstep} together with a reference that isolates
what the common cavity does.

\paragraph{What the common cavity changes.}  Noise suppression by a
rate-limited pump cycle is a single-emitter property, and the counting
statistics of independent sources superpose: for \(N\) independent
stationary sources whose output photons are counted together ---
summed count streams, as from \(N\) cavities on one detector or
\(N\) detectors whose counts are added, not coherent recombination of
the fields into one detected mode --- the long-time Fano factor of the
combined count is the rate-weighted mean of the individual ones, and
the equal-time correlation of the summed intensities is
\(g^{(2)}_{\rm tot}(0)=[g^{(2)}_1(0)+N-1]/N\).  The natural reference for
\(N\) emitters in one cavity is therefore \(N\) copies of the
single-emitter laser of the same scheme, each in its own cavity with the
same \(A\) and \(g\), driven so that the total photon number --- and
hence the mean output rate --- is the same, each at the smallest pump
rate giving \(\langle\nop\rangle_1=\langle\nop\rangle/N\), the same
rising-branch rule as in Table~\ref{tab:pumpstep} (open symbols in
Fig.~\ref{fig:pumpstep}; data in Ref.~\cite{code}).  In every row the
common cavity makes the output noisier than the independent reference.
For the \(\Lambda\) and ladder lasers the difference is small at
\(\langle\nop\rangle=0.3\) (\(F_{\rm count}=0.663\) against \(0.640\) for
the slowest ladder at \(N=2\); \(0.855\) against \(0.839\) for \(\Lambda\))
and appreciable at \(\langle\nop\rangle=1\) (\(0.740\) against \(0.553\) for
the \(\gamma_p=3A\) ladder at \(N=2\); \(0.820\) against \(0.658\) for
\(\Lambda\); \(0.858\) against \(0.709\) for \(\Lambda\) at \(N=3\)), where
the independent single-emitter lasers run at \(\langle\nop\rangle_1=1/N\)
each and keep more of their antibunching.  For the two-level and V
lasers the common cavity is what pushes the output above the shot-noise
level: their independent references are below one in every row
(\(0.93\)--\(0.96\) for the two-level laser) while the common-cavity
values exceed one in all but one row (\(0.97\)--\(1.31\)).  The equal-time statistics tell the same
story more strongly: the near-Poissonian \(g^{(2)}(0)\) of the
\(\Lambda\) and ladder lasers at \(\langle\nop\rangle=1\) (\(0.97\)--\(1.04\))
is a common-cavity effect, the independent reference being
\(0.85\)--\(0.91\).  Thus the sub-Poissonian counting statistics of the
multilevel lasers is not evidence of a collective enhancement --- the
common cavity erodes a single-emitter effect, by about \(0.02\) at
\(\langle\nop\rangle=0.3\) and by \(0.15\)--\(0.20\) at \(\langle\nop\rangle=1\)
for the \(\Lambda\) and ladder cycles, so that between about half and
\(94\%\) of the single-emitter suppression \(1-F_{\rm count}\) survives
(\(49\%\) for \(\Lambda\) at \(N=3\), \(\langle\nop\rangle=1\); \(94\%\) for the
slowest ladder at \(N=2\), \(\langle\nop\rangle=0.3\)) --- but the counting
statistics reveals substantial suppression that is much less apparent
in the equal-time statistics.  The comparison is an operational one
between two geometries at equal output: besides the inter-emitter
correlations, the occupation of each cavity and with it the stimulated
dynamics differ between the two arrangements, and the difference should
not be read as a measure of the correlations alone.  (Comparing at
the same pump rate instead of the same total photon number changes the
independent reference values by at most \(0.02\) for the \(\Lambda\) and
ladder schemes; for the two-level and V schemes the independent lasers
then have \(1.1\)--\(1.9\) times the photon number.)

\begin{figure}[tbp]
\includegraphics[width=\columnwidth]{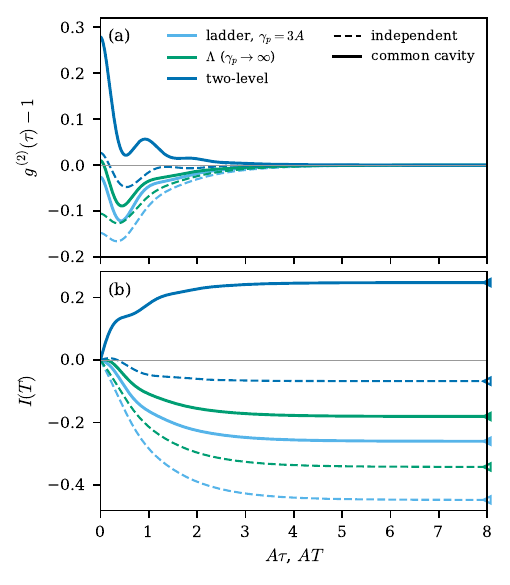}
\caption{Time-resolved origin of the counting-noise suppression, for
\(N=2\) emitters at \(\langle\nop\rangle=1\) and the pump rates of
Table~\ref{tab:pumpstep} (\(g=5A\), \(\gamma=A\), \(\gamma_r=3A\)): the
ladder with \(\gamma_p=3A\) (\(F_{\rm count}=0.740\)), its \(\Lambda\)
limit, and the two-level model.  (a) \(g^{(2)}(\tau)-1\); (b) cumulative
counting integral \(I(T)\), Eq.~(\ref{eq:IT}), whose limit is
\(F_{\rm count}-1\) (markers at the right edge: the Poisson-equation
values).  Solid: common cavity; dashed: the matched independent-source
reference (two single-emitter lasers at \(\langle\nop\rangle_1=1/2\),
summed count streams), for which \(g^{(2)}_{\rm tot}(\tau)-1=[g^{(2)}_1(\tau)-1]/2\)
and \(I_{\rm tot}=I_1\).  Computed by propagation with the truncated
\(K=0\) block-Jacobi matrix (\(n_{\max}=30\); curves change by less than
\(4\times10^{-14}\) at \(n_{\max}=40\)), not from residues.}
\label{fig:g2tau}
\end{figure}

\paragraph{Where the suppression sits in time.}  Integrated statistics
hide the time scale on which the regularity acts.
Figure~\ref{fig:g2tau} shows it for the headline case of
Table~\ref{tab:pumpstep}, two ladder emitters with \(\gamma_p=3A\) at
\(\langle\nop\rangle=1\), together with its \(\Lambda\) limit, the
two-level model, and the matched independent references, through
\(g^{(2)}(\tau)-1\) and the cumulative integral
\begin{equation}
I(T)=2A\langle\nop\rangle\int_0^T\big[g^{(2)}(\tau)-1\big]d\tau,\qquad
I(\infty)=F_{\rm count}-1 .
\label{eq:IT}
\end{equation}
Both follow from the \(K=0\) recurrence without residues:
\(g^{(2)}(\tau)-1\) is the trace functional of
\(e^{\mathcal M\tau}(Y-\langle\nop\rangle X_{\rm ss})\), and
\(I(T)=(2A/\langle\nop\rangle)\,\Tr[\nop\,(1-e^{\mathcal MT})Z]\) with the
Poisson solution \(Z\) above.  The anticorrelation is not at zero
delay.  For the ladder, \(g^{(2)}(\tau)\) falls from \(0.974\) at
\(\tau=0\) to a minimum of \(0.879\) at \(A\tau\approx0.43\); the
\(\Lambda\) laser is slightly bunched at zero delay (\(1.010\)) and has
its minimum, \(0.911\), at \(A\tau\approx0.45\).  The negative area
accumulates over two to three cavity lifetimes --- \(90\%\) of
\(I(\infty)\) is reached by \(AT\approx2.3\) for both lasers --- and
\(I(T)\) settles at \(-0.260\) and \(-0.181\), the values
\(F_{\rm count}-1\) of Table~\ref{tab:pumpstep}.  The minimum lies at a
delay of the order of, though shorter than, the mean time an emitter
needs to return from the lower to the upper laser level
(\(1/\gamma_r+1/P+1/\gamma_p\approx1/A\) for the ladder at this pump,
\(1/\gamma_r+1/P\approx1/A\) for \(\Lambda\)); since this time and the
cavity lifetime coincide here, Fig.~\ref{fig:g2tau} does not separate
the two.  The independent references have deeper minima that already
begin at \(\tau=0\) (\(g^{(2)}_{\rm tot}(0)=0.852\) and \(0.894\)), with
the minimum at a similar delay (\(A\tau\approx0.34\)--\(0.36\)): the common
cavity makes the dip shallower, removes most of it near zero delay, and
moves its minimum only slightly to longer delay; their \(I(T)\) settles
at the single-emitter values \(-0.447\) and \(-0.342\).  The two-level laser, by contrast,
has \(g^{(2)}(\tau)>1\) at all delays and accumulates a positive
\(I(\infty)=+0.248\), although its own independent reference has a
small negative area (\(-0.068\)).  This is why equal-time statistics
understate the effect of the pump cycle: \(g^{(2)}(0)\) samples the
correlation at the one delay at which the common-cavity dip has not yet
formed.

\paragraph{What the fitted two-level pump gets wrong.}  With the
parameters held fixed as stated above and only \(P\) fitted, the
two-level model reproduces \(\langle\nop\rangle\) by construction and
nothing else in Table~\ref{tab:pumpstep}: for \(N=2\) at
\(\langle\nop\rangle=1\) it gives \(g^{(2)}(0)=1.28\) and
\(F_{\rm count}=1.25\) against \(0.97\)--\(1.01\) and \(0.74\)--\(0.82\) for
the ladder family, and for \(N=2,3\) it places the output above the
shot-noise level in every row where all ladders and the \(\Lambda\)
scheme are below it.  At weak pump the fitted model also gets the order
and sign of the pair coherence wrong (Sec.~\ref{sec:weakpump}).
The reduction does become appropriate where one step of the cycle is
rate limiting and the others are fast compared with it and with the
cavity: the V scheme approaches the two-level model as
\(\gamma_r\to\infty\), with differences decreasing as \(1/\gamma_r\)
(Table~\ref{tab:Vconv}); the ladder with \(\gamma_p=100A\) is within
\(0.005\) of its \(\Lambda\) limit in \(F_{\rm count}\) in every row; and the
\(\Lambda\) scheme becomes the two-level model with pump rate
\(\gamma_r\), including its broadening \((\gamma+\gamma_r)/2\), when
\(P\gg\gamma_r\) (Sec.~\ref{sec:strongpump}).  The failures documented
here arise from finite recycling and pump-level bottlenecks, i.e.\ when
two or more steps of the cycle have comparable rates.  Whether a
two-level model with further fitted parameters could reproduce the
counting statistics as well we leave open.  How far the suppression can
be pushed by optimizing the cycle rates at given \(N\), and whether a
common cavity can ever improve on the independent reference, are
likewise open questions that the method can answer.

\section{Dark sectors}
\label{sec:dark}

For two-level emitters, Paper~I proved (its Proposition~8) that the collective coupling
annihilates the minimal collective-spin sector if and only if \(N\) is
even.  For \(d\ge3\), \(S_\pm\) act as the collective pseudo-spin ladder
of the emitters in laser levels and as zero on the others.  Let \(n_a\)
be the number of emitters in auxiliary levels and \(J\) the total
pseudo-spin of the remaining \(M=N-n_a\); the projectors
\(\Pi_{n_a,J}\) commute with permutations and sum to the identity, and
for pooled auxiliary levels the sector has dimension
\(\binom{N}{n_a}(d-2)^{n_a}(2J+1)m(M,J)\), with
\(m(M,J)=\binom{M}{M/2-J}-\binom{M}{M/2-J-1}\).  A sector is annihilated
by both \(S_\pm\) exactly when \(J=0\), so for every \(N\ge2\) --- and,
unlike for two-level emitters, also for odd \(N\) --- there are
interaction-dark sectors with active emitters (none is decoherence
free: local pump, recycling, and decay move population between them).
The sector-resolved feeding rate of Paper~I,
\(\Phi_{n_a,J}=-g\operatorname{Im}\langle a^\dagger S_-\Pi_{n_a,J}\rangle/\langle\Pi_{n_a,J}\rangle\),
connects the sectors to the output: by the first line of
Eq.~(\ref{eq:flux}),
\begin{equation}
A(\langle\nop\rangle-\nu)=\sum_{n_a,J}\langle\Pi_{n_a,J}\rangle\,\Phi_{n_a,J},
\label{eq:sectorflux}
\end{equation}
an exact decomposition of the emitted photon flux by sector, to which
the \(J=0\) sectors contribute nothing.

Table~\ref{tab:sectors} applies it to two to five \(\Lambda\) emitters
in the bad-cavity regime at \(P=A\).  The interaction-dark sectors with
active emitters hold \(10.8\%\), \(9.7\%\), \(8.3\%\), and \(6.6\%\) of
the population for \(N=2,\ldots,5\), slowly decreasing with no even/odd
alternation.  The output is carried disproportionately by the
high-pseudo-spin sectors, whose feeding rates grow faster than linearly
in \(J\) and depend little on \(N\), \(n_a\), and \(P\): at \(N=2\) the
bright triplet holds \(38.5\%\) of the population and emits \(70\%\) of
the photons, and at \(N=5\) the sectors with \(J\ge\tfrac32\) hold
\(43\%\) and emit \(73\%\).  How much more light the emitters would give
without the dark sectors is a counterfactual that requires their
residence times, which we have not computed.

\begin{table}[tbp]
\caption{Sector-resolved stationary state of \(N=2,\ldots,5\) \(\Lambda\)
emitters at \(g=1.1A\), \(\gamma=0.3A\), \(\gamma_r=A\), \(P=A\):
dimension, population (\%), feeding rate \(\Phi_{n_a,J}/A\), and share
of the output photon flux (\%), Eq.~(\ref{eq:sectorflux}), of the
sectors with \(n_a\) emitters in the reservoir and pseudo-spin \(J\) of
the rest.  Sectors with \(J=0\) are interaction dark.  The flux shares
add up to one to \(2\times10^{-15}\).}
\label{tab:sectors}
\begin{ruledtabular}
\begin{tabular}{ccccrrr}
\(N\) & \(n_a\) & \(J\) & dim & pop.\ (\%) & \(\Phi/A\) & flux (\%)\\
\hline
2 & 0 & $0$ & 1 & 10.8 & 0 & 0\\
2 & 0 & $1$ & 3 & 38.5 & 0.647 & 70.1\\
2 & 1 & $1/2$ & 4 & 41.7 & 0.255 & 29.9\\
2 & 2 & $0$ & 1 & 9.0 & 0 & 0\\
\hline
3 & 0 & $1/2$ & 4 & 15.2 & 0.252 & 7.1\\
3 & 0 & $3/2$ & 4 & 19.4 & 1.169 & 42.1\\
3 & 1 & $0$ & 3 & 9.7 & 0 & 0\\
3 & 1 & $1$ & 9 & 34.0 & 0.662 & 41.8\\
3 & 2 & $1/2$ & 6 & 18.9 & 0.259 & 9.1\\
3 & 3 & $0$ & 1 & 2.8 & 0 & 0\\
\hline
4 & 0 & $0$ & 2 & 2.3 & 0 & 0\\
4 & 0 & $1$ & 9 & 12.4 & 0.653 & 11.1\\
4 & 0 & $2$ & 5 & 9.4 & 1.796 & 23.3\\
4 & 1 & $1/2$ & 16 & 18.2 & 0.256 & 6.4\\
4 & 1 & $3/2$ & 16 & 22.6 & 1.200 & 37.4\\
4 & 2 & $0$ & 6 & 6.0 & 0 & 0\\
4 & 2 & $1$ & 18 & 20.4 & 0.674 & 19.0\\
4 & 3 & $1/2$ & 8 & 7.8 & 0.262 & 2.8\\
4 & 4 & $0$ & 1 & 0.9 & 0 & 0\\
\hline
5 & 0 & $1/2$ & 10 & 4.1 & 0.253 & 1.1\\
5 & 0 & $3/2$ & 16 & 8.3 & 1.184 & 10.7\\
5 & 0 & $5/2$ & 6 & 4.5 & 2.506 & 12.4\\
5 & 1 & $0$ & 10 & 3.6 & 0 & 0\\
5 & 1 & $1$ & 45 & 18.4 & 0.666 & 13.4\\
5 & 1 & $2$ & 25 & 13.6 & 1.847 & 27.4\\
5 & 2 & $1/2$ & 40 & 13.8 & 0.259 & 3.9\\
5 & 2 & $3/2$ & 40 & 16.9 & 1.224 & 22.5\\
5 & 3 & $0$ & 10 & 3.1 & 0 & 0\\
5 & 3 & $1$ & 30 & 10.4 & 0.682 & 7.8\\
5 & 4 & $1/2$ & 10 & 3.0 & 0.263 & 0.9\\
5 & 5 & $0$ & 1 & 0.3 & 0 & 0\\
\end{tabular}
\end{ruledtabular}
\end{table}

\section{Scaling with \texorpdfstring{\(N\)}{N} and the cumulant closure}
\label{sec:Nscaling}

\begin{figure*}[tbp]
\includegraphics[width=\textwidth]{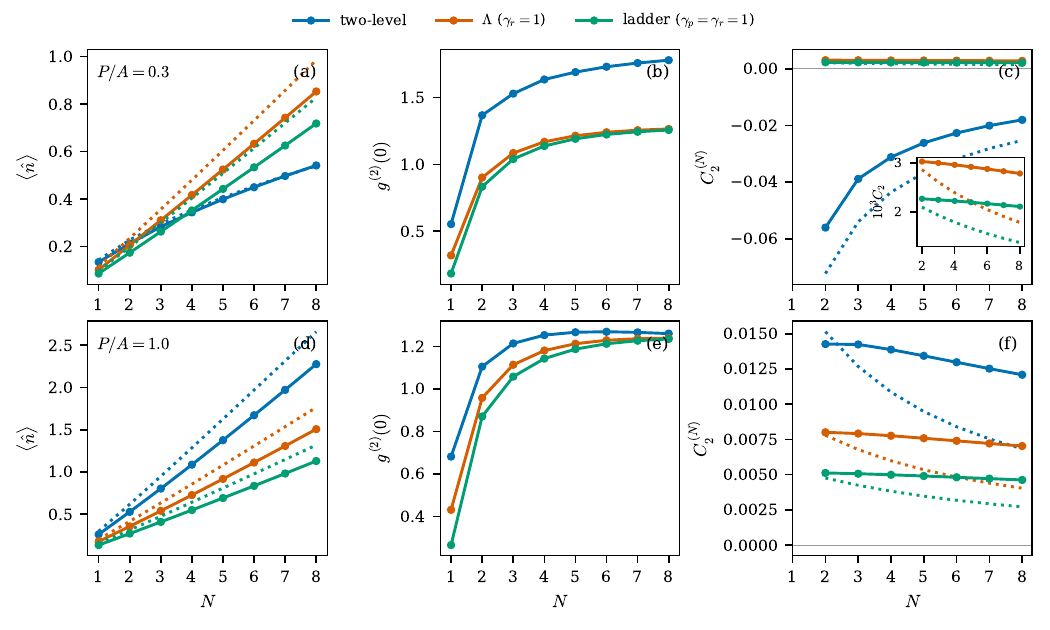}
\caption{Scaling with emitter number at \(g=1.1A\), \(\gamma=0.3A\),
\(\gamma_r=\gamma_p=A\), for \(P=0.3A\) (top) and \(P=A\) (bottom): (a,d)
photon number, (b,e) \(g^{(2)}(0)\), (c,f) pair coherence, for the
two-level, \(\Lambda\), and ladder schemes; the inset of (c) shows the
\(\Lambda\) and ladder coherences on their own scale.  Symbols: exact
continued fraction; dotted: second-order cumulant closure,
Appendix~\ref{app:closure}.  The \(N=8\) values are reproducible under a
change of radial cutoff from \(n_{\max}=30\) to \(40\) to better than
\(2\times10^{-13}\) (relative).}
\label{fig:Nscaling}
\end{figure*}

Figure~\ref{fig:Nscaling} shows \(\langle\nop\rangle\), \(g^{(2)}(0)\), and
\(C_2^{(N)}\) for \(N=1,\ldots,8\) in the bad-cavity regime at two pump
rates, for the two-level, \(\Lambda\), and ladder schemes, together with
the second-order cumulant closure of Appendix~\ref{app:closure}, the
multilevel form of the closure benchmarked in Paper~I.  For all three
schemes the photon number grows nearly linearly with \(N\) (per emitter:
\(0.11\), \(0.09\), and a saturating \(0.07\) for the \(\Lambda\), ladder,
and two-level schemes at \(P=0.3A\); \(0.18\), \(0.14\), \(0.28\) at
\(P=A\)).  Single emitters are strongly antibunched
(\(g^{(2)}(0)=0.32\), \(0.18\), \(0.55\) at \(P=0.3A\)); pairs of
\(\Lambda\) and ladder emitters are still sub-Poissonian (\(0.90\),
\(0.83\) at \(P=0.3A\); \(0.96\), \(0.87\) at \(P=A\)) where two-level pairs
are already bunched (\(1.37\), \(1.11\)), and all three schemes approach
\(g^{(2)}(0)\approx1.24\)--\(1.26\) by \(N=8\) at \(P=A\), while the
two-level laser at \(P=0.3A\) heads towards thermal statistics
(\(1.78\)).  The pair coherence of the \(\Lambda\) and ladder media is
positive and almost independent of \(N\) (\(+0.0030\) to \(+0.0028\) for
\(\Lambda\) at \(P=0.3A\)); the two-level coherence is negative at
\(P=0.3A\), decreasing in magnitude roughly as \(1/N\) as in Paper~I,
and positive at \(P=A\).

The closure is semi-quantitative for the photon number and poor for the
emitter variables.  It overestimates \(\langle\nop\rangle\) of the
\(\Lambda\) and ladder lasers by \(13\)--\(15\%\) at \(N=1\) rising to
\(15\)--\(17\%\) at \(N=8\) at both pump rates (two-level: \(+7\%\)
falling to \(0\%\) at \(P=0.3A\), \(+14\)--\(18\%\) at \(P=A\)).  It
underestimates the laser-level inversion by a factor of two to three
and a half for the \(\Lambda\) and ladder schemes at every \(N\), already
at \(N=1\) (\(0.036\) against the exact \(0.067\) for one \(\Lambda\)
emitter at \(P=0.3A\)), and by up to a factor of four for two-level
emitters at \(P=A\): the factorization
\(\langle\nop W_{ul}\rangle\to\langle\nop\rangle\langle W_{ul}\rangle\) misses
the anticorrelation of photon number and inversion identified in
Paper~I, which is stronger when the lower laser level is emptied by
recycling.  The sign of the pair coherence is correct in every case
shown, but its magnitude is underestimated by up to \(36\%\) (\(P=0.3A\))
and \(43\%\) (\(P=A\)) at \(N=8\) for the \(\Lambda\) and ladder media,
and overestimated by \(40\%\) for two-level emitters at \(P=0.3A\).  As in
Paper~I, the closure does not determine \(g^{(2)}(0)\) and therefore
cannot see the sub-Poissonian statistics of the multilevel pairs.

\begin{figure}[tbp]
\includegraphics[width=\columnwidth]{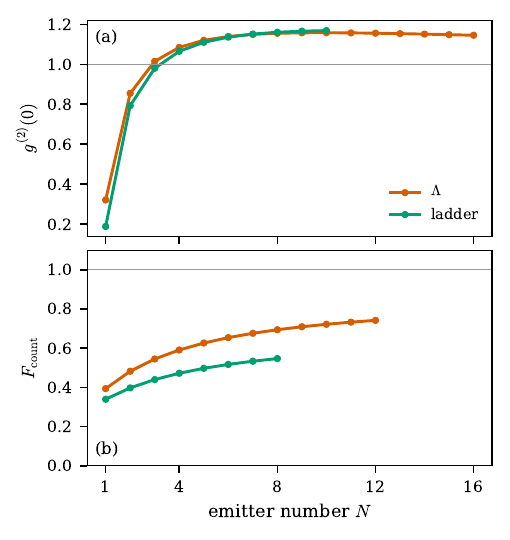}
\caption{(a) \(g^{(2)}(0)\) and (b) long-time counting Fano factor
versus emitter number for \(\Lambda\) and ladder emitters at the
benchmark point of Table~\ref{tab:timing} (\(g=1.1A\), \(\gamma=0.1A\),
\(P=0.5A\), \(\gamma_r=\gamma_p=A\); not the \(\gamma=0.3A\) series of
Fig.~\ref{fig:Nscaling}), from the norm-rescaled solver of
Appendix~\ref{app:normalized}.  Thin lines mark unity.}
\label{fig:largeN}
\end{figure}

\paragraph{Counting statistics at larger \(N\).}  Does the separation
between equal-time and counting statistics found in
Sec.~\ref{sec:statistics} persist as emitters are added?
Figure~\ref{fig:largeN} answers this at one benchmark point
(\(g=1.1A\), \(\gamma=0.1A\), \(P=0.5A\), \(\gamma_r=\gamma_p=A\)), using
the norm-rescaled solver of Appendix~\ref{app:normalized} beyond the
range of the plain recursion.  It does: although \(g^{(2)}(0)>1\) for
\(N\ge3\) (with a flat maximum of \(1.158\) near \(N=10\) for
\(\Lambda\) emitters), the long-time counting Fano factor stays below one
and rises only slowly with \(N\), from \(0.39\) at \(N=1\) to \(0.74\) at
\(N=12\) for \(\Lambda\) emitters and from \(0.34\) to \(0.55\) at \(N=8\)
for the ladder.  The regularity imposed by the rate-limited cycle thus
survives in the integrated output where the intracavity distribution is
already super-Poissonian, and is eroded only gradually by the common
cavity.  These are finite-\(N\) results at fixed single-emitter coupling
and fixed pump rate per emitter, so that the collective coupling grows
as \(\sqrt N\) and the total injection as \(N\); they are not a
large-\(N\) limit.  The stationary observables at this point, and their
reproducibility up to \(N=16\), are listed in Table~\ref{tab:timing}.

\section{Discussion and scope}
\label{sec:discussion}

\paragraph{Relation to other methods.}  Permutation-invariant
Fock-space solvers \cite{ChaseGeremia2008,XuTieriHolland2013,GeggRichter2016,Shammah2018}
work in the same reduced emitter space when the pumping is incoherent
\cite{GeggRichter2016} and give spectra and intensity correlations by
quantum regression, as our independent check of the counting statistics
does; the damping basis adds the block-Jacobi structure with its
provably unique linear closure and convergence theorem, the moment
functionals, the exactly terminating weak-pump series that produced
Table~\ref{tab:weakpump} and Fig.~\ref{fig:signmap}, and the regression
quantities within the same recurrence; the Fock solvers have no \(e^{\langle\nop\rangle}\) conditioning limit and
remain preferable above \(\langle\nop\rangle\approx20\)
(Appendix~\ref{app:validation}).  Relative to the one-atom multilevel
laser theory of Briegel, Meyer, and Englert \cite{Briegel1996}, which
derives an effective pump operator for one atom, the present
construction keeps the full collective structure of \(N\) emitters and
the full level kinetics.  The physical statements made here are
statements about the specified cycles at the specified rates, except
for Propositions~\ref{prop:pumpquantum} and \ref{prop:quench}, which
hold for any independent incoherent rate graph satisfying their
hypotheses; Proposition~\ref{prop:quench} is a sufficient condition for
quenching, and the shelving example of Sec.~\ref{sec:strongpump} shows
that it is not necessary.

\paragraph{Limitations.}  The construction requires identical emitters
with identical couplings and rates, one cavity-coupled transition, and
incoherent pumping; coherent pump fields, as in Raman-pumped single-ion
lasers \cite{Dubin2010}, enlarge the local space towards \(d^2\) and
change the grade bound (Remark~\ref{rem:general}).  Within these
restrictions, the differences between schemes found here matter where
two or more incoherent steps of the excitation cycle have comparable
rates, as for multistep carrier capture and relaxation in quantum-dot
gain media \cite{Gies2007}; the model parameters used here are
illustrative and not fitted to a specific device.  The double-precision
reach of the plain recursion in the damping basis is \(N\lesssim10\) for
three-level and \(N\lesssim8\) for four-level emitters; the
norm-rescaled coordinates of Appendix~\ref{app:normalized}
extend the tested range at the benchmark point (validated in detail at
\(N=12\), cross-checked through \(N=16\) for \(\Lambda\) and \(N=10\) for ladder emitters), but have not been applied to the parameter
scans of Secs.~\ref{sec:weakpump}--\ref{sec:statistics}, and the
\(e^{\langle\nop\rangle}\) limit \(\langle\nop\rangle\lesssim20\)--\(25\)
of the damping basis remains (Appendix~\ref{app:validation}).
The weak-pump orders are statements about the pump series of the
finite truncations (which have cutoff-independent coefficients); the
direct stationary bound transfers only to schemes in which every
charged level has a charge-lowering jump.  The convergence results and Proposition~\ref{prop:quench} are
for a zero-temperature cavity, the former under (U) and (P).  The dark-sector
populations are reported for \(N\le5\).  The matched-photon-number
pump-cycle comparisons of Sec.~\ref{sec:statistics} are for \(N\le3\);
the additional benchmark counting calculations of Sec.~\ref{sec:Nscaling}
reach \(N=12\) for \(\Lambda\) and \(N=8\) for ladder emitters, and the
stationary low-order observables \(N=16\) and \(N=10\), at one parameter
point.  The test of the effective two-level pump fits one parameter;
two-level models with further fitted parameters were not examined.
Open problems are the residence times of the dark sectors,
the optimization of pump-cycle rates for noise suppression at given
\(N\), a controlled enlargement of the local space for coherent drives,
and an analytic form of the sign boundary of Fig.~\ref{fig:signmap}.

\paragraph{Larger emitter numbers.}  With the norm-rescaled
coordinates the practical limitation is dense-block cost,
\(O(D_N^3)\) per recursion step with \(D_N=\binom{N+d+1}{d+1}\).  Sparse or
matrix-free solution of the stationary and Poisson problems, and, at
weak pump, a basis restricted by the filtration of
Proposition~\ref{prop:pumpquantum}, are natural routes beyond the low
tens; neither is implemented here.

\section{Conclusion}
\label{sec:conclusion}

The exact damping-basis continued fraction of Paper~I extends to
\(d\)-level emitters with arbitrary independent incoherent kinetics, in
the reduced operator space of populations and lasing coherences, with
the structural and convergence results intact under explicit
hypotheses.  The pump scheme is not a detail of few-emitter laser
modelling: pumping out of a laser level forces self-quenching, while the
\(\Lambda\) and ladder cycles studied here saturate to explicit limiting
models and convert a larger fraction of the absorbed excitation current
into cavity photons;
the number of pump quanta needed to enter the laser manifold sets the
order of the inter-emitter coherence, whose weak-pump sign is fixed by a
recycling-rate boundary; rate-limited steps in the pump cycle lower the
output-counting Fano factor of two and three emitters below one, a
single-emitter effect that the common cavity erodes but that remains
substantial in the counting statistics where the equal-time statistics
hide it, because the anticorrelation forms at finite delay; a two-level
pump fitted to the same photon number reproduces neither statistic for
two and three emitters, nor the order and sign of the pair coherence,
and becomes adequate where a single step of the cycle is rate limiting; and
reservoir levels create dark sectors at every emitter number, which
the exact sector decomposition of the output flux shows to emit
nothing while the high-pseudo-spin sectors carry most of the light.  Each of these is computed exactly at
the few-emitter numbers where cumulant closures, as the benchmark of
Sec.~\ref{sec:Nscaling} shows, are least reliable.

\section*{Data availability}
The code and data that produce every number, table, and figure of this
paper, including the regression tests and the independent Fock-space
checks, are openly available in the Zenodo repository of
Ref.~\cite{code} (doi:\codeDOI, version 1.0).  Its README maps every
result to a script; the principal ones are, in the folder
\texttt{results/}: Table~\ref{tab:weakpump}, Fig.~\ref{fig:signmap},
and the weak-pump sign boundary,
\texttt{fig2\_signmap\_weak\_pump\_series.py}, with the finite-pump
boundaries in \texttt{checks\_v3.py}; Table~\ref{tab:pumpstep},
\texttt{pump\_step\_comparison.py}; the independent-source comparison
of Sec.~\ref{sec:statistics}, \texttt{independent\_source\_reference.py};
Fig.~\ref{fig:g2tau}, \texttt{fig8\_g2\_tau.py}; the independent
Fock-space check of \(F_{\rm count}\),
\texttt{independent\_counting\_check.py}; the larger-\(N\) results of
Fig.~\ref{fig:largeN} and Table~\ref{tab:timing},
\texttt{large\_N\_validation.py}; and the stand-alone implementation of
Appendix~\ref{app:assembly}, \texttt{reconstruct\_from\_appendix.py}.

\begin{acknowledgments}
Claude Fable 5 (Anthropic) was used as an assistive tool for literature searches, manuscript editing, code generation, and symbolic and numerical checks, including the preparation of plots. All AI-assisted code and calculations were independently checked, and all references and text were independently reviewed by the author, who takes full responsibility for the manuscript.
\end{acknowledgments}

\appendix

\section{Construction details}
\label{app:construction}

\paragraph{Proof of Proposition~\ref{prop:invariant}.}
(i) For \(J=|i'\rangle\langle i|\),
\(\mathcal D[J]P_k=\delta_{ik}(P_{i'}-P_i)\) and
\(\mathcal D[J]\tau_+=J\tau_+J^\dagger-\tfrac12\{P_i,\tau_+\}=-\tfrac12(\delta_{iu}+\delta_{il})\tau_+\),
because \(\langle i|u\rangle\langle l|i\rangle=0\) for every \(i\);
likewise for \(\tau_-\).  \(\mathcal D[P_u]P_k=0\),
\(\mathcal D[P_u]\tau_\pm=-\tfrac12\tau_\pm\), and the level Hamiltonian
commutes with every \(P_k\) and gives
\([P_i,\tau_\pm]=\pm(\delta_{iu}-\delta_{il})\tau_\pm\).  (ii)
\(\tau_+P_k=\delta_{kl}\tau_+\), \(P_k\tau_+=\delta_{ku}\tau_+\),
\(\tau_+\tau_-=P_u\), \(\tau_-\tau_+=P_l\), \(\tau_\pm^2=0\), and the
mirror identities.  (iii) \(\openone=\sum_kP_k\).  The Liouvillian is a
sum of one-emitter Liouvillians, field terms, and the interaction, which
acts on the emitters only by left and right multiplication with
\(\tau_\pm^{(j)}\); so \(\mathcal F\otimes\mathcal A^{\otimes N}\) is
invariant, and so is its permutation-invariant part.  The maximally
mixed emitter state lies in this closed subspace, and for a Fock
truncation its evolution converges to the unique stationary state, which
therefore lies in the subspace; the convergence of Fock truncations
(Appendix~\ref{app:carry}) transfers the statement to \(\rho_{\rm ss}\).
The dimension is the number of weak compositions of \(N\) into \(d+2\) parts.

A coherent drive on any transition, or a second cavity-coupled
transition, enlarges \(\mathcal A\) by the coherences it generates, up to
the full \(d^2\); Remark~\ref{rem:general} says what survives.

\begin{table}[tbp]
\caption{Atomic block dimension \(\binom{N+L-1}{L-1}\) for the reduced
local space, \(L=d+2\), and for the full local Liouville space,
\(L=d^2\).}
\label{tab:dims}
\begin{ruledtabular}
\begin{tabular}{rrrrrr}
\(N\) & \(d=2\) & \(d=3\) & \(d=4\) & \(d=3\), full & \(d=4\), full\\
\hline
 2 & 10 & 15 & 21 & 45 & 136\\
 3 & 20 & 35 & 56 & 165 & 816\\
 4 & 35 & 70 & 126 & 495 & 3876\\
 6 & 84 & 210 & 462 & 3003 & 54264\\
 8 & 165 & 495 & 1287 & 12870 & 490314\\
12 & 455 & 1820 & 6188 & 125970 & \(1.7\times10^{7}\)\\
\end{tabular}
\end{ruledtabular}
\end{table}

\paragraph{Local bases.}
Two bases of \(\mathcal A\) are used in the physics sections, both with exactly one traceful
element \(b_0\) and traceless \(b_1,\ldots,b_{d-1}\), so that the
normalization and the observable functionals below are the same in
either.  The \emph{damping basis} consists of
\(r_\kappa=\sum_i(w_\kappa)_iP_i\), with \(w_0=\bar p\) and
\(w_1,\ldots,w_{d-1}\) the remaining right eigenvectors of \(W\)
(traceless, because the left null vector of \(W\) is \((1,\ldots,1)\)),
and \(r_\pm=\tau_\pm\); these are eigenoperators of the one-emitter
Liouvillian with eigenvalues \(-\beta_\kappa\) (the eigenvalues of
\(-W\), complex for cyclic rate graphs) and \(-C_{ul}\mp i\Omega\),
Eq.~(\ref{eq:Cul}).  The \emph{fixed basis} consists of \(P_{g}\) for a
reference level \(g\), the traceless \(P_i-P_g\) (\(i\neq g\)), and
\(\tau_\pm\); it is independent of all rates, the atomic part of the
recurrence is not diagonal in it, and it is the basis in which the pump
series of Sec.~\ref{sec:weakpump} is constructed.  The damping basis is
used when \(W\) is diagonalizable with a well-conditioned eigenvector
matrix, the fixed basis otherwise (defective \(W\) do occur, e.g.\ the
\(\Lambda\) scheme at \(P=\gamma=\gamma_r/4\), with eigenvalues
\(0,-3\gamma_r/4,-3\gamma_r/4\)); the implementation \cite{code} switches
automatically and the two give identical results where both apply
(Appendix~\ref{app:validation}).  The cavity damping basis
\(\rho_n^{(k)}\) consists of the right eigenoperators of
\(\mathcal L_{\rm f}=A(1+\nu)\mathcal D[a]+A\nu\mathcal D[a^\dagger]-i\omega[\nop,\cdot]\);
left and right multiplication by \(a\), \(a^\dagger\) change \(k\) by one
and \(n\) by at most one through the coefficients
\(f^{\rm L/R}_{\mu,\delta}(n,k)\) of Table~\ref{tab:fieldcoeff},
Appendix~\ref{app:assembly}.
A third, rate-independent orthogonal basis is used for the largest
emitter numbers (Appendix~\ref{app:normalized}).

\paragraph{Proof of Lemma~\ref{lem:lift}.}
Each of the \(\mathcal N(\mv)\) arrangements in \(R_{\mv}\) has \(m_\mu\)
sites carrying \(b_\mu\); replacing one of them by \(b_\nu\) yields an
arrangement of type \(\mv'=\mv-e_\mu+e_\nu\), and each such arrangement
is produced \(\mathcal N(\mv)m_\mu/\mathcal N(\mv')=m'_\nu\) times, so
that the sum over sites and arrangements is
\(x_{\nu\mu}m_\mu\mathcal N(\mv)R_{\mv'}\).

Applied to \(\tau_+\cdot\), with \(\tau_+r_\kappa=(w_\kappa)_l\tau_+\) and
\(\tau_+\tau_-=P_u=\sum_\kappa c^{(u)}_\kappa r_\kappa\), the rule gives
\(S_+R_{\mv}=\sum_\kappa m_\kappa(w_\kappa)_lR_{\mv-e_\kappa+e_+}+m_-\sum_\kappa c^{(u)}_\kappa R_{\mv-e_-+e_\kappa}\)
and three analogous formulas, which for \(d=2\) reproduce Eqs.~(6)--(9)
of Paper~I term by term; the same rule lifts the one-emitter
Liouvillian.  In the implementation the structure constants are obtained
by projecting \(\tau_\pm b_\mu\), \(b_\mu\tau_\pm\), and
\(\mathcal L_{\rm loc}b_\mu\) onto \(\mathcal A\), the vanishing of the
projection residual being the numerical form of
Proposition~\ref{prop:invariant}.

\begin{remark}[Scope]
\label{rem:general}
Items (i)--(ii) and the nilpotency in (iii) hold for any finite
collection of local operators closed under the one-emitter Liouvillian
and under multiplication by the cavity-coupled operators, provided a
level charge exists that every term of \(S_+\) raises by exactly one and
that the coherent part of the dynamics conserves.  The grade range, and
with it the nilpotency index and the length of the Neumann series, is
then \(|K|+N(h_{\max}-h_{\min})\) with \(h\) the spectral range of the
local charge, not \(N+|K|\): a three-level ladder with both adjacent
transitions coupled to the same mode has charges \(0,1,2\) and index up
to \(2N+|K|+1\).  We do not pursue such models here.
\end{remark}

\paragraph{Numerical solution.}
For \(N\le10\) we extract \(M^{(1)},M^{(0)},F^{(1)},G\) by finite
differences at a bulk index, set \(R_{n_{\max}}=R_\infty\), recurse down
to \(R_0\), take \(X_0\) as the right singular vector of \(M_0+GR_0\) for
its smallest singular value, normalize the component along
\(\Phi_{0,Ne_0}\) to one, and reconstruct \(X_{n+1}=R_nX_n\); for larger \(N\) the norm-rescaled
form of Appendix~\ref{app:normalized} is used.

\section{What the inherited proofs use}
\label{app:carry}

We list, for each item of Proposition~\ref{prop:carry}, the properties
of the emitter sector on which the corresponding proof in Paper~I
rests, and why they hold for Eq.~(\ref{eq:master}).

(i) The uncoupled atomic Liouvillian is diagonal (damping basis) or a
general matrix (fixed basis) on the \(R_{\mv}\), which the recurrence does
not need to be diagonal; the interaction
changes \(n\) by at most one through the field identities, which are
untouched; and \(K\) is conserved because \(\tau_+\) raises \(q\) by
exactly one on every nonvanishing term of the lifted multiplication ---
\(\tau_+r_\kappa\propto\tau_+\) and \(\tau_+\tau_-\in\operatorname{span}\{r_\kappa\}\)
--- and \(\tau_-\) lowers it.

(ii) The proof of Proposition~2 of Paper~I identifies \(M^{(1)}\) from
the \(n\)-linear field coefficients alone and then uses only that
\([S_+,\cdot]\) raises and \([S_-,\cdot]\) lowers the charge of every
basis state by one, together with \(|q-K|\le N+|K|\).

(iii) Proposition~3 of Paper~I uses the field coefficients and the same
charge property for \(F^{(1)}\), \(\mathcal N_K\), and \(G\); the
statement for \(R_\infty\) is that each term of its Neumann series
lowers the grade by \(j+1\ge1\), so that \(R_\infty\) lowers it by at
least one (the proof of Proposition~3 of Paper~I says ``at least'';
grade changes of \(-1\) and \(-2\) both occur, e.g.\ for two \(\Lambda\)
emitters).

(iv) Proposition~4 of Paper~I uses only the affine structure of the
blocks, constant \(G\), the invertibility of \(M^{(1)}\), and, for the
extension from \(n\ge n_0\) to all \(n\), hypothesis (P), stated there
in the same form.

(v) In the proof of Lemma~1 of Paper~I, \(H\) commutes with \(\Exc\)
because \(a^\dagger S_-\) and \(aS_+\) do; the jumps \(a\) and
\(|i'\rangle\langle u|\) lower \(\Exc\) by one, the jumps
\(|u\rangle\langle i|\) raise it by one at total rate
\(\sum_j\sum_{i\neq u}\gamma_{u\leftarrow i}P_i^{(j)}\le\Gamma_\uparrow N\),
and all other jumps preserve it, so
\(\mathcal L^\dagger e^{\theta\Exc}=e^{\theta\Exc}Q_\theta\) (called \(W\) there) with
\(Q_\theta\le(e^{-\theta}-1)A\nop+(e^\theta-1)\Gamma_\uparrow N\) at \(\nu=0\), and
the rest of the proof is unchanged.  The proof of Proposition~5 of
Paper~I uses this bound, the compactness of the resolvent of \(\Exc\),
and (U); the proof of Theorem~1 (Appendix~C of Paper~I) uses the bound,
item~(iv), the grade bound \(|k_\alpha|\le N\), the \(U(1)\) covariance
of the master equation --- which holds for Eq.~(\ref{eq:master}) under
\(|u\rangle\to e^{i\varphi}|u\rangle\), \(a\to e^{i\varphi}a\), since
\(\mathcal D[|i'\rangle\langle i|]\) is invariant under any rephasing of
the levels --- and permutation invariance.  None of these refer to
\(d=2\).  The normalization convention (component along
\(\Phi_{0,Ne_0}\) equal to one) requires a local basis with a single
traceful element, which both bases of Sec.~\ref{sec:method} provide.

\section{Second-order cumulant closure for multilevel emitters}
\label{app:closure}

The exact equations of motion for \(\langle\nop\rangle\), the one-emitter
level populations \(p_i=\langle P_i^{(j)}\rangle\),
\(Z_1=\langle a^\dagger\tau_-^{(j)}\rangle\), and
\(C_2=\langle\tau_+^{(i)}\tau_-^{(j)}\rangle\) (\(i\neq j\)) couple to the
third-order moments \(\langle a^\dagger a\,(P_u-P_l)^{(j)}\rangle\) and
\(\langle a^\dagger(P_u-P_l)^{(i)}\tau_-^{(j)}\rangle\).  Setting the
corresponding third-order cumulants to zero, as in Eq.~(33) of Paper~I
(the first moments of the charged operators vanish by \(U(1)\)
symmetry), gives, at \(\nu=0\) and with \(w=p_u-p_l\),
\begin{align}
\frac{d}{dt}\langle\nop\rangle&=-A\langle\nop\rangle-gN\operatorname{Im}Z_1,\notag\\
\frac{d}{dt}p&=Wp+g\operatorname{Im}Z_1\,(e_u-e_l),\notag\\
\frac{d}{dt}Z_1&=-(A/2+C_{ul}+i\Delta)Z_1\notag\\&\quad-\tfrac{ig}{2}\big[p_u+(N-1)C_2+\langle\nop\rangle w\big],\notag\\
\frac{d}{dt}C_2&=-2C_{ul}C_2-g\,w\operatorname{Im}Z_1 ,
\label{eq:closure}
\end{align}
with \(W\) the rate matrix (\ref{eq:W}).  For \(d=2\) these are the
closure equations of Paper~I, and the implementation reproduces its
quoted closure values (\(C_2=-0.017\), \(\langle\sigma_z\rangle_{\rm cl}=-0.039\)
at \(N=3\), \(s=0.9\), baseline parameters).  The physical root is
obtained by integrating Eq.~(\ref{eq:closure}) from the vacuum with all
emitters in the stationary distribution of \(W\).  The closure contains
no independent fourth-order field moment and does not determine
\(g^{(2)}(0)\).

\section{Validation, cost, conditioning}
\label{app:validation}

\paragraph{Validation.}  (a) With \(d=2\) the implementation
\cite{code} reproduces the two-level code of Paper~I: the weak-pump
reference value \(\langle\nop\rangle/s=0.34521131203\) of its Appendix~D\,2
to \(10^{-11}\) (relative) and the dominant
spectral poles of its Table~III to all nine digits shown (both are
regression tests of the package).  (b) For the V, \(\Lambda\), cyclic \(\Lambda\)
(with pure dephasing), and ladder schemes at \(N=1,2\),
\((\Delta,\nu)\in\{(0,0),(0.4A,0.1)\}\), and for the cyclic \(\Lambda\)
scheme at \(N=3\), the photon number, \(g^{(2)}(0)\), all level
populations, \(Z\), the pair coherence, and every element of the
reconstructed reduced atomic state agree with a brute-force
steady-state solution of the full \(d^N\times n_{\rm cav}\) master
equation (no symmetry reduction, \(n_{\rm cav}=16\), or 12 at \(N=3\))
to \(10^{-15}\)--\(10^{-13}\) at \(\nu=0\) and to \(5\times10^{-11}\) at
\(\nu=0.1\), the latter limited by the brute-force cutoff.  (c) The
largest-photon-number point of the scan of Fig.~\ref{fig:pumpscan}
(\(N=3\), two-level, \(P/A=32\)) agrees with the brute-force solution at Fock cutoffs 70 and
85 to all eight digits of \(\langle\nop\rangle=24.925794\),
\(g^{(2)}(0)=1.0114823\), and \(C_2=0.10702979\), so that the
representation limit discussed below does not affect it.  (d) At the
defective-\(W\) point \(P=\gamma=\gamma_r/4\) of the \(\Lambda\) scheme
the automatic fallback to the fixed basis reproduces the raw-projector
basis and the brute force to \(10^{-15}\) at \(N=1,2\), and the damping,
fixed, and raw bases agree to \(2\times10^{-14}\) at neighboring points
(recorded in the validation output; the regression tests assert
\(10^{-12}\) and \(10^{-11}\)).  (e) The
structural statements of Proposition~\ref{prop:carry} were checked for
all four schemes at \(N\le3\), \(K\in\{0,-1,2\}\), \(\nu=0.1\),
\(\Delta=0.3A\): the deviations of the explicitly constructed \(M_n\),
\(G_n\), \(F_n\) at \(n=0,1,13\) from their affine forms are at most
\(3\times10^{-13}\), every eigenvalue of \(\mathcal N_K\) vanishes to
\(3\times10^{-15}\), the nilpotency index of \(\mathcal N_K\) and of
\(R_\infty\) equals \(N+|K|+1\) exactly at \(K=0\) and \(K=-1\) in every
case, and every nonzero element of \(F^{(1)}\), \(\mathcal N_K\), and
\(G\) connects grades differing by exactly one in the stated direction.
(f) The V scheme converges to the two-level model as
\(\gamma_r\to\infty\) at the expected rate (Table~\ref{tab:Vconv}): at
\(N=2\), \(P=8A\), the differences in all four observables decrease as
\(1/\gamma_r\), which is the quantitative form of the statement that
the two-level incoherent pump is the fast-reservoir limit of the V
scheme, including its pump-induced broadening \(C=(\gamma+P)/2\).  (g)
The dominant pole of the regression spectrum is selected as the pole
with the largest real part among those whose residue modulus (a
conjugate pair counting jointly) exceeds a threshold; for all schemes
of Fig.~\ref{fig:pumpscan} at \(P/A=3,16,64\) the selected width is
identical for thresholds of \(1\%\), \(5\%\), and \(20\%\).  (h) The
counting Fano factor was recomputed independently in the full
Fock\(\otimes\)emitter space for five cases of Table~\ref{tab:pumpstep}
(two-level \(N=1\); \(\Lambda\) \(N=2\); ladder \(N=2\) at
\(\langle\nop\rangle=0.3\) and \(1\); ladder \(N=3\)), by building the
Liouvillian with QuTiP and solving the Poisson equation for the
connected regression integral; at Fock cutoffs 10--13 (11 at \(N=3\))
the results agree with the continued fraction to \(6\times10^{-15}\),
\(2\times10^{-14}\), \(2\times10^{-13}\), \(9\times10^{-13}\), and
\(5\times10^{-13}\).  (i) Strong-pump limits and finite-pump
boundaries (Secs.~\ref{sec:weakpump} and \ref{sec:strongpump}; data in
\texttt{results/checks\_v2} and \texttt{results/checks\_v3} of
Ref.~\cite{code}).  At \(N=2\), \(g=5A\), \(\gamma=A\), the \(\Lambda\)
laser with \(\gamma_r=10A\) has \(\langle\nop\rangle=6.167\), \(6.623\),
\(6.672\), \(6.677\) at \(P/A=64\), \(640\), \(6400\), \(64000\) (limit
\(6.678\)), the fast ladder \(4.113\), \(4.318\), \(4.339\), \(4.342\)
(V limit \(4.342\)), the two-level laser \(0.084\), \(0.0079\),
\(0.00078\) and the V laser \(0.080\), \(0.0075\), \(0.00074\) at
\(P/A=640\), \(6400\), \(64000\); the shelving cycle has
\(\langle\nop\rangle=0.0675\), \(0.0198\), \(0.00245\), \(0.000251\) at
\(P/A=1\), \(10\), \(100\), \(1000\).  The pair coherence at \(P=16A\)
vanishes at \(\gamma_r/A=3.545\) (\(\Lambda\)) and \(3.511\) (ladder) for
\(N=2\) and at \(3.253\) and \(3.243\) for \(N=3\), against \(3.547\) for
both schemes and both \(N\) at \(P=10^{-3}A\) and the weak-pump series
value \(3.547\).  (j) The norm-rescaled solver used for \(N>10\) and its validation are
described in Appendix~\ref{app:normalized}.
Checks (b), (c), and (h) are the comparisons with
independent solvers; the series coefficients, sign map, spectra, and
\(N\le8\) results are validated through the structural checks and
internal consistency, not individually by brute force.

\begin{table}[tbp]
\caption{Convergence of the V scheme to the two-level incoherent-pump
model as the reservoir decay \(\gamma_r\to\infty\): differences of the
photon number, \(g^{(2)}(0)\), pair coherence, and dominant-pole width
\(\Gamma_1=-2\operatorname{Re}\lambda_1\) from the two-level values at
\(N=2\), \(g=5A\), \(\gamma=A\), \(P=8A\) (two-level:
\(\langle\nop\rangle=5.4665\), \(g^{(2)}(0)=1.0447\), \(C_2=0.0403\),
\(\Gamma_1=0.303A\)).}
\label{tab:Vconv}
\begin{ruledtabular}
\begin{tabular}{rrrrr}
\(\gamma_r/A\) & \(\delta\langle\nop\rangle\) & \(10^3\delta g^{(2)}\) & \(10^3\delta C_2\) & \(10^3\delta\Gamma_1/A\)\\
\hline
20 & $-0.953$ & $-14.8$ & $-7.62$ & $+22.3$\\
80 & $-0.273$ & $-4.63$ & $-2.21$ & $+5.47$\\
320 & $-0.0708$ & $-1.23$ & $-0.576$ & $+1.36$\\
1280 & $-0.0179$ & $-0.312$ & $-0.146$ & $+0.339$\\
5120 & $-0.00448$ & $-0.0782$ & $-0.0366$ & $+0.0848$\\
20480 & $-0.00112$ & $-0.0196$ & $-0.00915$ & $+0.0212$\\
\end{tabular}
\end{ruledtabular}
\end{table}

\begin{table*}[!htb]
\caption{Block dimension, time of one stationary solve at \(n_{\max}=30\),
and reproducibility, for \(\Lambda\) emitters (\(d=3\)) and ladder
emitters (\(d=4\)) at \(g=1.1A\), \(\gamma=0.1A\), \(P=0.5A\),
\(\gamma_r=\gamma_p=A\).  ``Plain'': the recursion in the damping basis
used for Secs.~\ref{sec:weakpump}--\ref{sec:Nscaling}, with the largest
relative change of \(\langle\nop\rangle\) and \(g^{(2)}(0)\) between
\(n_{\max}=30\) and \(40\) (at \(N=12\) it fails, see text).
``Norm-rescaled'': the solver of Appendix~\ref{app:normalized},
whose values are listed; ``cutoff'' is the largest relative change of
\(\langle\nop\rangle\), \(g^{(2)}(0)\), \(C_2\) between \(n_{\max}=30\)
and \(40\) (\(^b\)between \(20\) and \(30\)), ``basis'' the same against
the fixed local basis (\(^a\)with grade scaling \(\lambda=0.3\)).  The
spreads are reproducibility measures, not accuracy bounds.}
\label{tab:timing}
\begin{ruledtabular}
\begin{tabular}{ccrrrcrcccll}
 & & & & \multicolumn{2}{c}{plain} & \multicolumn{6}{c}{norm-rescaled}\\
\(d\) & \(N\) & \(D_N\) & \(D_N\) (full) & time (s) & spread & time (s) & \(\langle\nop\rangle\) & \(g^{(2)}(0)\) & \(C_2^{(N)}\) & cutoff & basis\\
\hline
3 & 4 & 70 & 495 & 0.03 & $9\times10^{-15}$ & 0.02 & 0.73539 & 1.08490 & $+0.01356$ & $<10^{-15}$ & ---\\
3 & 6 & 210 & 3003 & 0.3 & $5\times10^{-13}$ & 0.1 & 1.13318 & 1.13941 & $+0.01214$ & $<10^{-15}$ & $2\times10^{-15}$\\
3 & 8 & 495 & 12870 & 2.5 & $2\times10^{-10}$ & 0.9 & 1.54136 & 1.15534 & $+0.01102$ & $<10^{-15}$ & $7\times10^{-14}$\\
3 & 10 & 1001 & 43758 & 15 & $3\times10^{-7}$ & 5.1 & 1.95685 & 1.15823 & $+0.01010$ & $<10^{-15}$ & $3\times10^{-12}$\\
3 & 12 & 1820 & 125970 & 82 & $7\times10^{-3}$ & 26 & 2.37779 & 1.15586 & $+0.00934$ & $10^{-13}$ & $3\times10^{-10}$\\
3 & 14 & 3060 & 319770 & --- & --- & 130 & 2.80299 & 1.15121 & $+0.00869$ & $2\times10^{-12}$\,$^b$ & $4\times10^{-12}$\,$^a$\\
3 & 16 & 4845 & 735471 & --- & --- & 424 & 3.23162 & 1.14560 & $+0.00813$ & $3\times10^{-11}$\,$^b$ & $2\times10^{-10}$\,$^a$\\
4 & 4 & 126 & 3876 & 0.1 & $<10^{-15}$ & 0.05 & 0.59504 & 1.06512 & $+0.01004$ & $<10^{-15}$ & ---\\
4 & 6 & 462 & 54264 & 2.0 & $4\times10^{-14}$ & 0.7 & 0.91285 & 1.13596 & $+0.00910$ & $<10^{-15}$ & $3\times10^{-15}$\\
4 & 8 & 1287 & 490314 & 31 & $5\times10^{-13}$ & 9.5 & 1.23819 & 1.16061 & $+0.00834$ & $<10^{-15}$ & $3\times10^{-14}$\\
4 & 10 & 3003 & 3268760 & --- & --- & 120 & 1.56896 & 1.16877 & $+0.00770$ & $10^{-15}$\,$^b$ & $2\times10^{-12}$\\
\end{tabular}
\end{ruledtabular}
\end{table*}

\paragraph{Cost.}  The blocks are affine in \(n\) and are generated from
four stored \(D_N\times D_N\) matrices.  The plain stationary solver
stores all transfer matrices \(R_n\) for the upward reconstruction, so
its memory is \(O(n_{\max}D_N^2)\): \(0.8\)~GB for eight ladder emitters
(\(D=1287\), \(n_{\max}=30\)).  The norm-rescaled solver of Appendix~\ref{app:normalized}
streams the recursion and keeps only \(R_0\) and \(R_1\), so its memory
is \(O(D_N^2)\) (a few dense blocks, about \(3\)~GB at \(D=4845\)).  The
time of both is \(O(n_{\max}D_N^3)\).  The spectrum and the intensity
correlation diagonalize a dense matrix of dimension \((n_{\max}+1)D_N\);
the block Poisson solver of Appendix~\ref{app:normalized} gives the counting Fano factor
without an eigendecomposition at the cost of one stationary solve with
stored \(R_n\).  The dense routes are reported separately; they dominate the cost of Figs.~\ref{fig:pumpscan}
and \ref{fig:recycling} (about four minutes for the \(N=3\) ladder
curve at 16 pump values).  Table~\ref{tab:timing} lists stationary-solve
timings on a two-core virtual machine (Intel Xeon, 2.8~GHz) with numpy~2.5
linked to OpenBLAS running two threads, for both solvers.

\paragraph{Conditioning.}  The precision limits are those of Paper~I:
the \(e^{\langle\nop\rangle}\) representation limit of the damping basis
in double precision --- which check (c) shows not to affect the
\(\langle\nop\rangle\approx25\) point at \(N=3\), where the blocks are
small --- and the \(N\)-dependent conditioning of the pivots, which comes
from the non-normality of the nilpotent part of \(M^{(1)}\) (index
\(N+1\), as before) and sets in at smaller \(N\) because the blocks are
larger.  In Table~\ref{tab:timing} the plain double-precision recursion
in the damping basis is reproducible under a change of radial cutoff to \(10^{-12}\)
(\(6\times10^{-12}\) for \(C_2\)) for eight ladder emitters, to \(10^{-10}\) (\(5\times10^{-10}\) for \(C_2\))
for eight \(\Lambda\) emitters, and to \(3\times10^{-7}\) (\(10^{-6}\) for
\(C_2\)) for ten; for twelve \(\Lambda\) emitters \(\langle\nop\rangle\)
and \(g^{(2)}(0)\) change by \(0.7\%\) and \(0.8\%\) between
\(n_{\max}=30\) and \(40\), the pair coherence is not reproducible at
all, and the photon number is \(20\%\) off, because in these
coordinates the stationary kernel of \(M_0+GR_0\) is not separated from
the next singular values (Appendix~\ref{app:normalized}).  The practical reach of the plain
recursion is thus \(N\lesssim10\) for three-level and \(N\lesssim8\) for
four-level emitters.  The norm-rescaled solver extends this range at
the benchmark point: its results are reproducible under a change of radial
cutoff to \(10^{-13}\) (relative) for \(D_N\le2000\)
(\(n_{\max}=30\) versus \(40\)) and to \(2\times10^{-12}\) and
\(3\times10^{-11}\) at \(N=14\) and \(16\) (\(20\) versus \(30\)); they agree with the fixed local basis
to \(10^{-12}\) for ladder emitters through \(N=10\), and with the
grade-scaled fixed basis (\(\lambda=0.3\)) to \(4\times10^{-12}\) at
\(N=14\) and \(2\times10^{-10}\) at \(N=16\) for \(\Lambda\) emitters
(without grade scaling the fixed basis itself degrades with \(N\), to
\(4\times10^{-6}\) at \(N=16\)); at \(N=12\) they agree with an
extended-precision recursion to \(3\times10^{-14}\).  These are
reproducibility measures across independent representations, not
rigorous error bounds.
Where the full local space would be needed (coherent drives) the reach
at comparable block sizes is \(N\le5\) for \(d=3\) and \(N\le3\) for
\(d=4\).

\section{Norm-rescaled solver for larger \texorpdfstring{\(N\)}{N}}
\label{app:normalized}

\paragraph{Why the plain recursion fails.}  The reach of the plain
recursion (Table~\ref{tab:timing}, ``plain'' columns) is limited by the
coordinates, not by the equations.  In the damping basis the three
smallest singular values of \(M_0+GR_0\) at \(N=12\) (\(\Lambda\)
benchmark of Table~\ref{tab:timing}) are \(1.7\times10^{-7}\),
\(3.0\times10^{-7}\), \(3.6\times10^{-7}\): the stationary kernel is not
separated, and the singular vector used as kernel gives a photon number
\(20\%\) below the value obtained below.

\paragraph{Coordinates and blocks.}  The norm-rescaled solver
(implementation \cite{code}, \texttt{core/normalized.py}) solves the same
recurrence after three exact changes of representation.  (1) The local
basis is \(b_0=\openone/d\), \(b_1,\ldots,b_{d-1}\) Hilbert--Schmidt
orthonormal traceless diagonal operators, and \(b_\pm=\tau_\pm\).
(2) The collective basis is rescaled to unit norm: for mutually
orthogonal \(b_\mu\),
\(\|R_{\mv}\|_{\rm HS}^2=\prod_\mu\|b_\mu\|^{2m_\mu}/\mathcal N(\mv)\), and
the new coordinates are \(x_{\mv}=X_{\mv}\|R_{\mv}\|_{\rm HS}\), so that all
blocks undergo the diagonal similarity
\(B\to\mathcal W^{-1}B\mathcal W\), \(\mathcal W=\operatorname{diag}(\|R_{\mv}\|_{\rm HS}^{-1})\).
The normalization and observable prescriptions of Sec.~\ref{sec:method}
must be transformed accordingly.  Only \(R_{Ne_0}=(\openone/d)^{\otimes N}\)
is traceful, with \(\Tr R_{Ne_0}=1\) and
\(\|R_{Ne_0}\|_{\rm HS}=d^{-N/2}\); the trace condition
\(X_{0,Ne_0}=1\) therefore becomes \(x_{0,Ne_0}=d^{-N/2}\), not one, and an
observable functional \(v_{\mv}=\Tr(\mathcal O R_{\mv})\) becomes
\(v_{\mv}/\|R_{\mv}\|_{\rm HS}\).  (An overall constant factor in
\(\mathcal W\) is immaterial; the implementation fixes it for numerical
range and converts back to \(X_{\mv}\) before evaluating observables.)
(3) The affine coefficients \(M^{(1)}\), \(M^{(0)}\), \(G\), \(F^{(1)}\) are
assembled from the scalar field coefficients and sparse lifted one-site
maps, rather than by differencing blocks at two radial indices, so that
\(G^{(1)}=F^{(0)}=0\) hold exactly rather than to rounding.

\paragraph{Algorithms.}  The downward recursion is streamed and keeps
only \(R_0\) and \(R_1\), which suffice for \(X_0,X_1,X_2\) and hence for
\(\langle\nop\rangle\), \(g^{(2)}(0)\), the populations, and \(C_2\) (memory
\(O(D_N^2)\)).  The kernel is obtained by replacing one linearly
dependent row of \(M_0+GR_0\) (the row of the traceful element, which
carries the trace functional as a left null vector) by the trace
condition, instead of a singular-value decomposition.  The counting Fano
factor is obtained in the same coordinates from the Poisson equation of
Sec.~\ref{sec:statistics}, \(\mathcal M Z=-(Y-\langle\nop\rangle X_{\rm ss})\),
by the block analogue of the continued fraction:
\(Z_{n+1}=R_nZ_n+t_{n+1}\) with the stationary \(R_n\), \(t_{n_{\max}}=0\),
\(t_n=(M_n+GR_n)^{-1}(b_n-Gt_{n+1})\), and at \(n=0\) the same row
replacement with \(\Tr Z_0=0\).  Only \(Z_0\), \(Z_1\) enter the
integral, but the source needs all \(X_n\), so the \(R_n\) are stored
(memory \(O(n_{\max}D_N^2)\)); no eigendecomposition of the
\((n_{\max}+1)D_N\)-dimensional matrix is needed.

\paragraph{Validation.}  At \(N=12\) the solver gives
\(\langle\nop\rangle=2.3777888932\), \(g^{(2)}(0)=1.1558613090\),
\(C_2=0.0093404530\).  These values are reproduced under four
variations of the numerical representation (not independent physical
solvers): radial cutoffs \(20\), \(30\), \(40\) (relative spread
\(2\times10^{-14}\), \(10^{-13}\) for \(C_2\)); additional grade rescalings
\(\lambda^{|k|}\), \(\lambda=0.3\) and \(3\) (\(10^{-13}\) and
\(5\times10^{-11}\)); the fixed local basis (\(P_{\rm ref}\),
\(P_i-P_{\rm ref}\)), non-orthogonal and related to the first by a dense
transformation of every block (\(6\times10^{-11}\); \(10^{-12}\) with
\(\lambda=0.3\)); and the recursion and kernel solve repeated in extended
precision (64-bit mantissa, Gaussian elimination with partial pivoting)
on the same double-precision blocks (\(5\times10^{-15}\)).  The last check
tests the propagation of rounding through the elimination, not the
assembly of the blocks.  Consistency diagnostics of the solution are
recorded in the repository: at \(N=12\), \(14\), \(16\) (\(\Lambda\)) and
\(N=8\), \(10\) (ladder) the normwise relative residual of the \(n=0\) and
\(n=1\) block equations is below \(3\times10^{-18}\) (that of the full
truncated system at \(N=12\) and \(8\) likewise), the trace condition and
the sum of the level populations hold to \(2\times10^{-16}\), all
populations lie in \([0,1]\), the imaginary parts of \(\langle\nop\rangle\),
\(\langle\nop^2\rangle\), and \(C_2\) vanish, and the \(\Lambda\) populations
satisfy the stationary flux balance \(Pp_r=\gamma_rp_l\) to
\(5\times10^{-16}\).  Small residuals do not by themselves bound the
error of an ill-conditioned problem, which is why the variations above
are the primary evidence.  The counting solver
reproduces the dense Poisson solve and the independent Fock value
\(0.7399491367\) of Appendix~\ref{app:validation}, check (h), to
\(2\times10^{-14}\), is cutoff independent to \(4\times10^{-15}\) up to
\(N=10\), and agrees with the fixed basis to \(7\times10^{-12}\) at
\(N=12\).  The spreads of the same variations through \(N=16\)
(\(\Lambda\)) and \(N=10\) (ladder) are given in Table~\ref{tab:timing} and
Appendix~\ref{app:validation} (``Conditioning'').  All spreads are
reproducibility measures, not accuracy bounds.

\paragraph{Ablation.}  Relative deviation of \(\langle\nop\rangle\) at
\(N=12\), \(n_{\max}=30\), from the reference calculation (orthogonal
basis, norm rescaling, algebraic blocks): plain solver as in
Table~\ref{tab:timing}, \(0.2\); same blocks with the trace-row kernel,
\(1\times10^{-4}\); trace-row kernel without norm rescaling,
\(3\times10^{-4}\) (damping basis), \(2\times10^{-9}\) (fixed), and
\(2\times10^{-12}\) (orthogonal); with norm rescaling but differenced
blocks, \(4\times10^{-14}\) (orthogonal).  The kernel extraction and the
local basis therefore matter most; the norm rescaling contributes the
last two orders.  The damping basis is not rescued by the norm
rescaling (\(5\times10^{-5}\), and \(3\%\) with \(\lambda=3\)): its
eigenvector populations are far from orthogonal.

\section{Assembly of the recurrence}
\label{app:assembly}

This appendix lists everything needed to implement the recurrence of
Sec.~\ref{sec:method} without Paper~I.  A stand-alone implementation
that follows it line by line and imports nothing from the package,
\texttt{results/reconstruct\_from\_appendix.py} of Ref.~\cite{code}, reproduces
the package to \(2\times10^{-14}\) (relative) for the \(N=2\) rows of
Table~\ref{tab:pumpstep} at \(\langle\nop\rangle=1\) and for ladder
emitters at \(\nu=0.1\), \(\Delta=0.4A\).  We work in the frame rotating
at the cavity frequency \(\omega\), in which the term \(-i\omega K\) of
Sec.~\ref{sec:cf} is absent.

\paragraph{Cavity damping basis.}  With \(k_\pm=\max(\pm k,0)\), the
generalized Laguerre polynomials \(L_n^{(|k|)}\), and \(:\;:\) normal
ordering,
\begin{equation}
\rho_n^{(k)}=\frac{(-1)^n}{(\nu+1)^{|k|+1}}\,(a^\dagger)^{k_+}
:\!L_n^{(|k|)}\!\Big(\frac{\nop}{\nu+1}\Big)e^{-\nop/(\nu+1)}\!:\,a^{k_-}
\label{eq:rhonk}
\end{equation}
for \(n\ge0\), \(k\in\mathbb Z\) \cite{BriegelEnglert1993}; it satisfies
\(\mathcal L_{\rm f}\rho_n^{(k)}=-A(n+|k|/2)\rho_n^{(k)}\) and
\(\Tr\rho_n^{(k)}=\delta_{n0}\delta_{k0}\).  (At \(\nu=0\),
\(:\!f(\nop)e^{-\nop}\!:\) is evaluated with
\(:\!\nop^je^{-\nop}\!:=(a^\dagger)^j|0\rangle\langle0|a^j\).)  Writing
\(a^{(-1)}=a\), \(a^{(+1)}=a^\dagger\), the field multiplications are
\begin{align}
a^{(\mu)}\rho_n^{(k)}&=\sum_{\delta}f^{\rm L}_{\mu,\delta}(n,k)\,\rho_{n+\delta}^{(k+\mu)},\notag\\
\rho_n^{(k)}a^{(\mu)}&=\sum_{\delta}f^{\rm R}_{\mu,\delta}(n,k)\,\rho_{n+\delta}^{(k+\mu)},
\label{eq:fieldmult}
\end{align}
\(\delta\in\{-1,0,1\}\), \(\rho_{-1}^{(k)}=0\), with the nonzero
coefficients of Table~\ref{tab:fieldcoeff}.  Equations~(\ref{eq:rhonk})
and (\ref{eq:fieldmult}) and the table were checked against each other
in a Fock space of dimension \(70\) for \(n\le5\), \(|k|\le3\),
\(\nu=0\) and \(0.3\), to \(2\times10^{-15}\)
(\texttt{results/check\_field\_coefficients.py}).

\begin{table}[tbp]
\caption{Nonzero field coefficients \(f^{\rm L/R}_{\mu,\delta}(n,k)\) of
Eq.~(\ref{eq:fieldmult}), with \(r=\nu/(1+\nu)\); all others vanish.}
\label{tab:fieldcoeff}
\begin{ruledtabular}
\begin{tabular}{llccc}
operation & range & \(\delta=-1\) & \(\delta=0\) & \(\delta=+1\)\\
\hline
\(a\rho_n^{(k)}\) & \(k>0\) & & \(n+k\) & \((n+1)r\)\\
 & \(k\le0\) & \(1+\nu\) & \(\nu\) & \\
\(a^\dagger\rho_n^{(k)}\) & \(k\ge0\) & \(1+\nu\) & \(1+\nu\) & \\
 & \(k<0\) & & \(n+|k|\) & \(n+1\)\\
\(\rho_n^{(k)}a\) & \(k>0\) & & \(n+k\) & \(n+1\)\\
 & \(k\le0\) & \(1+\nu\) & \(1+\nu\) & \\
\(\rho_n^{(k)}a^\dagger\) & \(k\ge0\) & \(1+\nu\) & \(\nu\) & \\
 & \(k<0\) & & \(n+|k|\) & \((n+1)r\)\\
\end{tabular}
\end{ruledtabular}
\end{table}

\paragraph{Local structure constants.}  Represent an element of
\(\mathcal A\) by its coordinates \((p_1,\ldots,p_d,c_+,c_-)\) in
\(\{P_1,\ldots,P_d,\tau_+,\tau_-\}\).  By the identities in the proof of
Proposition~\ref{prop:invariant}, the five one-site maps act as
\begin{align}
\tau_+x&:\ (c_+',p',c_-')=(p_l,\,c_-e_u,\,0),\notag\\
x\tau_+&:\ (c_+',p',c_-')=(p_u,\,c_-e_l,\,0),\notag\\
\tau_-x&:\ (c_-',p',c_+')=(p_u,\,c_+e_l,\,0),\notag\\
x\tau_-&:\ (c_-',p',c_+')=(p_l,\,c_+e_u,\,0),\notag\\
\mathcal L_{\rm loc}x&:\ p'=Wp,\quad c_\pm'=-(C_{ul}\pm i\Delta)\,c_\pm,
\label{eq:localmaps}
\end{align}
with \(W\) of Eq.~(\ref{eq:W}) and \(C_{ul}\) of Eq.~(\ref{eq:Cul}).  For
a basis \(b_\mu=\sum_iB_{i\mu}\,(\text{coordinate vector } i)\), the
structure constants of a map with coordinate matrix \(x\) are
\(B^{-1}xB\).  We use the fixed basis \(b_0=P_g\),
\(b_i=P_i-P_g\) (\(i\neq g\)), \(b_\pm=\tau_\pm\), whose only traceful
element has \(\Tr b_0=1\), so that \(\Tr R_{\mv}=\delta_{\mv,Ne_0}\).
Lifting the five maps with Eq.~(\ref{eq:lift}) gives the
\(D_N\times D_N\) matrices \(S^{\rm L}_\pm\), \(S^{\rm R}_\pm\) (left and
right multiplication by \(S_\pm\)) and \(\Lambda^{\rm at}\) (the emitter
Liouvillian including detuning).  The charge of \(R_{\mv}\) is
\(q=m_+-m_-\), and in the sector \(K\) basis state \(\alpha\) carries
the field grade \(k_\alpha=K-q_\alpha\).

\paragraph{Blocks.}  With \(\mathcal L_{\rm int}\rho=(ig/2)[a^\dagger S_-+aS_+,\rho]\)
and the diagonal matrices
\(\mathsf f^{\rm L/R}_{\mu,\delta}(n)=\operatorname{diag}_\alpha f^{\rm L/R}_{\mu,\delta}(n,k_\alpha)\),
the coupling of \(X_n\) into the equation for \(X_{n+\delta}\) is
\begin{align}
B^{(\delta)}_n={}&\frac{ig}{2}\Big[S^{\rm L}_-\mathsf f^{\rm L}_{+,\delta}(n)
+S^{\rm L}_+\mathsf f^{\rm L}_{-,\delta}(n)\notag\\
&\quad-S^{\rm R}_-\mathsf f^{\rm R}_{+,\delta}(n)
-S^{\rm R}_+\mathsf f^{\rm R}_{-,\delta}(n)\Big],\notag\\
M_n={}&-A\operatorname{diag}_\alpha\big(n+|k_\alpha|/2\big)+\Lambda^{\rm at}+B^{(0)}_n,\notag\\
G_n={}&B^{(-1)}_{n+1},\qquad F_n=B^{(+1)}_{n-1},\qquad F_0=0 .
\label{eq:blocks}
\end{align}
The affine coefficients of Proposition~\ref{prop:carry} are
\(M^{(1)}=M_{n+1}-M_n\), \(M^{(0)}=M_0\), \(F^{(1)}=F_{n+1}-F_n\), and
\(G=G_n\).

\paragraph{Solution and observables.}  For \(K=0\): set
\(R_{n_{\max}}=R_\infty=-(M^{(1)})^{-1}F^{(1)}\), recurse
\(R_n=-(M_{n+1}+G_{n+1}R_{n+1})^{-1}F_{n+1}\) down to \(R_0\), solve
\((M_0+G_0R_0)X_0=0\) with the row of \(Ne_0\) replaced by the
normalization \(X_0[Ne_0]=1\), and set \(X_{n+1}=R_nX_n\).  Observables
follow from Sec.~\ref{sec:obs}.  For the counting statistics, the
coefficient vector \(Y\) of \(a\rho_{\rm ss}a^\dagger\) is obtained by
applying Eq.~(\ref{eq:fieldmult}) twice to every component
\((X_n)_\alpha\), first \(f^{\rm L}_{-,\delta_1}(n,k_\alpha)\) and then
\(f^{\rm R}_{+,\delta_2}(n+\delta_1,k_\alpha-1)\), the atomic index being
unchanged; \(F_{\rm count}\) then follows from the Poisson equation of
Sec.~\ref{sec:statistics} for the truncated block-tridiagonal matrix
\(\mathcal M\) built from Eq.~(\ref{eq:blocks}).  The emission spectrum
uses the same blocks with \(K=-1\).  As a checksum, two ladder emitters
with \(g=5A\), \(\gamma=A\), \(\gamma_p=\gamma_r=3A\), \(P=3A\),
\(\Delta=\nu=0\) give \(\langle\nop\rangle=0.99574218\),
\(g^{(2)}(0)=0.97279395\), and \(F_{\rm count}=0.73821790\)
(\(n_{\max}=30\) and \(40\) agree to \(10^{-14}\)).

\bibliography{refs}

\end{document}